\documentclass[lettersize,journal]{IEEEtran}
\usepackage{amsmath,amsfonts}
\usepackage{algorithmic}
\usepackage{array}
\usepackage{textcomp}
\usepackage{stfloats}
\usepackage{amsthm}
\usepackage{url}
\usepackage{amsmath}
\usepackage{verbatim}
\usepackage{algorithm,algorithmic}
\usepackage{xcolor}
\usepackage{tabularx}
\newtheorem{theorem}{Theorem}
\usepackage{subfigure}
\newtheorem{lemma}{Lemma}
\usepackage{graphicx}
\usepackage{cite}
\usepackage[hidelinks]{hyperref}
\def\BibTeX{{\rm B\kern-.05em{\sc i\kern-.025em b}\kern-.08em
    T\kern-.1667em\lower.7ex\hbox{E}\kern-.125emX}}
\usepackage{balance}
\usepackage[bottom]{footmisc}
\newcommand\omicron{o}
\usepackage{multicol}
\usepackage{physics}
\usepackage[bottom]{footmisc}
\usepackage{esint}
\DeclareMathOperator{\Div}{div}
\usepackage{mathtools}
\begin{document}

\title{Reconstructing classes of 3D FRI signals from sampled tomographic projections at unknown angles}

\author{Renke Wang, \textit{Student Member, IEEE},\ Francien G. Bossema,\ Thierry Blu, \textit{Fellow, IEEE},\\ and Pier Luigi Dragotti, \textit{Fellow, IEEE}

\thanks{The work in this paper was in part presented at EUSIPCO 2023~\cite{me2}.
}}

\maketitle
\begin{abstract}
Traditional sampling schemes often assume that the sampling locations are known. Motivated by the recent bioimaging technique known as cryogenic electron microscopy (cryoEM), we consider the problem of reconstructing an unknown 3D structure from samples of its 2D tomographic projections at unknown angles. We focus on 3D convex bilevel polyhedra and 3D point sources and show that the exact estimation of these 3D structures and of the projection angles can be achieved up to an orthogonal transformation. Moreover, we are able to show that the minimum number of projections needed to achieve perfect reconstruction is independent of the complexity of the signal model. By using the divergence theorem, we are able to retrieve the projected vertices of the polyhedron from the sampled tomographic projections, and then we show how to retrieve the 3D object and the projection angles from this information. The proof of our theorem is constructive and leads to a robust reconstruction algorithm, which we validate under various conditions. Finally, we apply aspects of the proposed framework to calibration of X-ray computed tomography (CT) data.
\end{abstract}

\begin{IEEEkeywords}
sampling, sampling at unknown locations, cryogenic electron microscopy (cryoEM), unknown view tomography (UVT), finite rate of innovation (FRI)
\end{IEEEkeywords}

%\onecolumn

\section{Introduction}

\IEEEPARstart{S}{ampling} plays an important role in signal processing. It refers to the mechanism which converts continuous signals into discrete sequences~\cite{unser_sampling-50_2000}. From the foundational Whittaker-Shannon theorem~\cite{shannon_communication_1949} to the latest advancements in compressive sensing~\cite{candes_robust_2006,donoho_compressed_2006}, finite rate of innovation~\cite{pina1, blu1, pld1}, and super resolution~\cite{candes_towards_2014,guo_super-resolving_2023}, sampling theories have provided perfect reconstruction conditions when the precise locations of the samples are known. 

However, in many applications, the sampling locations may not be available or only available partially. This happens, for example, in a classical problem in robotics where a location-unaware mobile sensor attempts to jointly determine its own locations and reconstruct a map using measurements from the environment (SLAM)~\cite{alexandru_diffusion_2021,guo_fri_2020}. Likewise in computer vision, one tries to reconstruct a 3D scene from a series of images from different unknown viewpoints (SfM)~\cite{hartley_multiple_2004}. In structural biology, the latest technique to reconstruct a high resolution 3D structure of particles of interest is cryogenic electron microscopy (cryoEM), where one has only access to parallel beam projections of the particle at unknown angles~\cite{bendory_single-particle_2020,cryoem2,basu,basu_journal}. In medical computed tomography (CT), projection angles are assumed to be known, however, they can be inaccurate due to patient motion or machine mis-calibration.

For the aforementioned reconstruction problems, it is not feasible to apply directly traditional sampling methods. Instead, we need to consider the problem of sampling at unknown locations. Compared to prior works, sampling at unknown locations has been so far less-explored. Marziliano \textit{et al.}\cite{pina2} considered the reconstruction of discrete-time bandlimited signals with unknown sampling locations. Another discrete-time formulation is unlabeled sensing, where a linear sensing system contains unlabeled observation data~\cite{martin2}. 
As for the continuous-time counterpart, Browning~\cite{browning} introduced an alternating least squares algorithm for bandlimited signals. Nordio \textit{et al.}~\cite{nordio} also provided asymptoptic performance analysis for the reconstruction of bandlimited signals using linear techniques. While Kumar~\cite{kumar1,kumar2} took a statistical perspective and considered the unknown sampling locations to be statistically dependent. He also showed that the reconstruction error is asymptotically inversely proportional to the number of samples. Recently, Elhami \textit{et al.}~\cite{martin1} showed that by constraining the sampling positions to belong to some known function spaces, it is possible to transform the location-unaware signal recovery problem to that of regular sampling of a composite of functions.

Despite the encouraging results, the theoretical works have so far only considered more traditional sampling setups where the samples are directly taken from the signal. Motivated by cryoEM, in this work, we consider a specific case for the location-unaware sampling, where the aim is to reconstruct an unknown 3D structure from samples of its 2D parallel beam tomographic projections at unknown projection angles.

Conventional methods directly retrieve the projection angles using common-line based algorithms~\cite{commonline}, then the 3D structure is estimated through filtered back-projection~\cite{backprojection1, backprojection2} or iterative regularized optimization~\cite{opt1,opt2}. Another category of methods alternate between the estimation of the projection angles and the refinement of the 3D structure. This can be achieved by projection matching~\cite{projmatch1, projmatch2,projmatch3,projmatch4,mona1} or formulating the 3D refinement step as a maximum marginalized \emph{a posterior} problem~\cite{bayesian1,bayesian2,bayesian3,bayesian4}. The third category of methods completely bypass the process of estimating the projection angles by either distribution matching~\cite{cryogan} or estimating rotational invariant features~\cite{invariant1, invariant2, invariant3}.

These methods often adopt a discrete perspective on the projection measurements or angles. While this simplification leads to practical implementation, it does not fully model the underlying continuous process. Moreover, there is still no theoretical analysis of sufficient conditions on e.g. the number of projections that ensures a perfect reconstruction. 

In this work, we focus on specific classes of signals with finite rate of innovation (FRI), specifically convex bilevel polyhedra and 3D point sources, and address the fundamental sampling question of when perfect reconstruction of the 3D structure can be achieved, given only samples of a limited number of 2D projections at unknown angles. We present a method with a constructive proof that allows the simultaneous estimation of the 3D signal and projection angles. The estimation is exact up to an orthogonal transformation. Moreover, we show that the minimum number of projections needed is three and is irrespective of the complexity of the 3D signal.

In addition to its theoretical contribution, the proposed method can be applied to the geometrical calibration of real CT systems. During the calibration process, radiographs of reference objects (usually spherical markers) at equidistant angles are often used to determine the system geometry and to correct the geometrical offsets \cite{pointmarker}. Nevertheless, in certain instances, acquiring equidistant radiographs may prove unattainable due to misalignment or the absence of advanced equipment. We show that the proposed method can be applied to the calibration of projection angles of real CT data obtained at the FleX-ray laboratory at the Center for Mathematics \& Computer Science, Amsterdam.

The rest of the manuscript is organized as follows. In Section~\ref{section2}, we define the signal model and
formulate the reconstruction problem from sampled projections at unknown angles. Then, in Section~\ref{section3}, we present an exact reconstruction method that achieves the simultaneous estimation of the 3D signal and the projection angles. In Section~\ref{section4}, we propose alternative strategies to robustify the method in case the samples are corrupted with noise. We then provide numerical results on synthetic data in Section~\ref{section5}, and show that the proposed framework can be applied for calibration in real computed tomography problems. Finally, we conclude in Section~\ref{section7}. 

\section{Problem formulation} \label{section2}

\begin{figure}[t]
\centering
\includegraphics[width=0.3\textwidth]{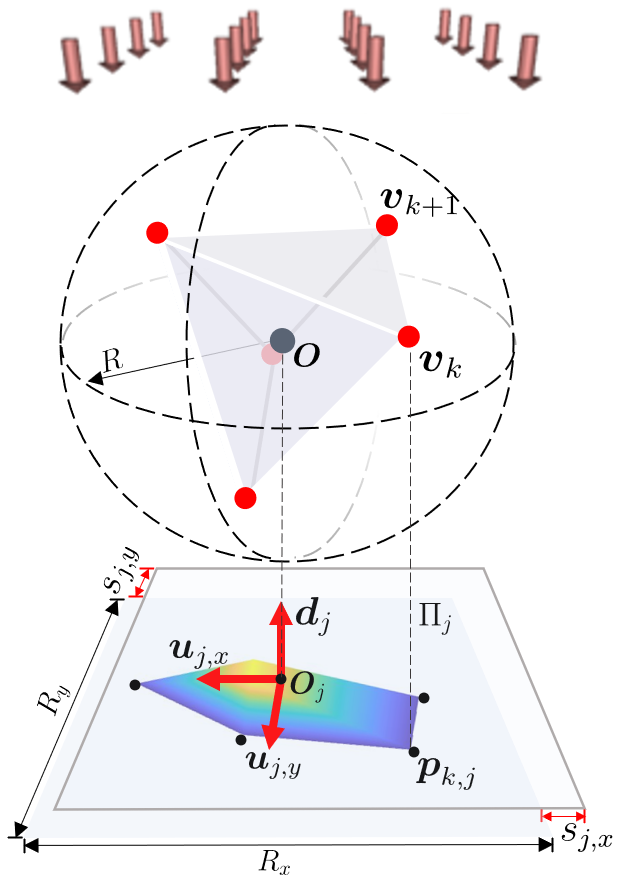}
\caption{An example of a bilevel convex polyhedron model $g(\mathbf{r})$ with $K=5$ vertices. The figure also depicts the parallel beam projection onto the observation window $\Pi_{j}$. Due to an unknown amount of shift, the center of the actual observation window (grey boundary) deviates from the projection of the origin $\boldsymbol{O}_{j}$ by $s_{j,x},\ s_{j,y}$ along the two edges $\boldsymbol{u}_{j,x}$, $\boldsymbol{u}_{j,y}$ respectively.}
\label{fig:short}
%\vspace*{-0.4cm}
\end{figure}

We consider a 3D convex, non-degenerate polyhedron with $K$  vertices $\{\boldsymbol{v}_{k}\}_{k=1}^{K}$, where $K\in\mathbb{N}$ and $K\geq 4$, $\boldsymbol{v}_{k}\in\mathbb{R}^{3}$ denotes the coordinate of the $k$th vertex. The origin is assumed to be the geometric center of the vertices: $\sum_{k=1}^{K}\boldsymbol{v}_{k}=\boldsymbol{0}$, and all vertices lie within a
sphere region of known radius $R\in\mathbb{R}^{+}$ centered at the origin, as shown in Fig.~\ref{fig:short}. 

The convex hull $\partial\Gamma$ of the vertices is then specified by : 
\begin{equation*}
    \partial\Gamma = \Biggl\{\sum_{k=1}^{K}\lambda_{k} \boldsymbol{v}_{k}: \forall \lambda_{k}\in\mathbb{R},\ \lambda_{k}\geq 0 \text{ and } \sum_{k=1}^{K}\lambda_{k} = 1\Biggr\}.
\end{equation*}

We define its corresponding 3D bilevel, convex and simply connected polyhedron model $g(\boldsymbol{r})$ as follows: 
\begin{equation} \label{eq1}
    g(\boldsymbol{r}) = 
\begin{cases}
1,\ \text{for } \boldsymbol{r} \in \Gamma,\\
0,\ \text{ otherwise},
\end{cases}
\forall \boldsymbol{r} \in \mathbb{R}^{3}
\end{equation}
where $\Gamma$ is the 3D volume enclosed by $\partial \Gamma$.

\begin{figure}[t]
\centering
\includegraphics[width=0.4\textwidth]{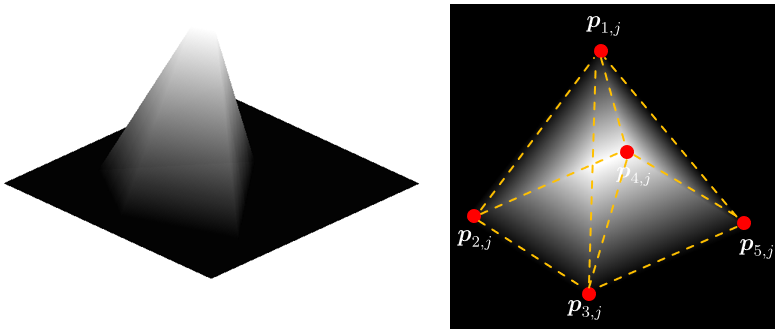}
\caption{An example of a tomographic projection $I_{j}(x,y)$ of the polyhedron shown in Fig.~1. The projection can be decomposed into 6 disjoint triangular regions with linear changes of intensity. The projected vertices $\boldsymbol{p}_{k,j}$ are denoted with red dots.}
\label{fig:example_projection}
%\vspace*{-0.4cm}
\end{figure}

Assume $J\geq 3$ parallel-beam projections of the 3D volume $g(\boldsymbol{r})$ are taken. The projection angles are assumed to be distinct, and are represented by unit direction vectors $\boldsymbol{d}_{j}\in\mathbb{R}^{3},\ j=1,...,J$. On the $j$th projected 2D plane, we consider a rectangular observation window $\Pi_{j}$ whose center $\boldsymbol{O}_{j}\in\mathbb{R}^{3}$ corresponds to the projection of the origin $\boldsymbol{O}$, as shown in Fig.~\ref{fig:short}. We also assume that the observation window $\Pi_{j}$ containing the 2D projection is of known dimension $(R_{x}, R_{y})$ such that $R_{x}, R_{y}> 2R$. The 2D tomographic projection $I_{j}(x,y)$ can then be expressed as the Radon transform of $g(\boldsymbol{r})$ onto the $j$th observation window $\Pi_{j}$ as follows:
\begin{align} \label{eq3}
     \nonumber I_{j}(x,y) &= \int_{\mathbb{R}^3}g(\boldsymbol{r}) \delta\left(x-\boldsymbol{r}^{T}\boldsymbol{u}_{j,x}-
    s_{j,x}\right)\\
    &\cdot\delta\left(y-\boldsymbol{r}^{T}\boldsymbol{u}_{j,y}-
    s_{j,y}\right)d^{3}\boldsymbol{r},\ j=1,...,J
\end{align}
where
\begin{itemize}
    \item $\boldsymbol{u}_{j,x}, \boldsymbol{u}_{j,y}\in\mathbb{R}^{3}$ denotes the unit direction vectors representing the two edges of the rectangular window $\Pi_{j}$, and they satisfy:
\begin{equation} \label{eq2}
\boldsymbol{u}_{j,x}\times\boldsymbol{u}_{j,y} = \boldsymbol{d}_{j},\ j = 1,...,J
\end{equation}
where $\times$ denotes vector product. Moreover, vectors in $\{\boldsymbol{u}_{j,x}\}_{j=1}^{J}$ are assumed to be pairwise distinct. The same assumption holds true for $\{\boldsymbol{u}_{j,y}\}_{j=1}^{J}$.
    \item $\boldsymbol{s}_{j}=[s_{j,x}$, $s_{j,y}]^{T}\in\mathbb{R}^{2}$ denotes the shifts of the observation window $\Pi_{j}$ along $\boldsymbol{u}_{j,x}$, $\boldsymbol{u}_{j,y}$ directions respectively. Thus, the center of the actual observation window is $  \boldsymbol{O}_{j}'=\boldsymbol{O}_{j}+s_{j,x}\boldsymbol{u}_{j,x}+s_{j,y}\boldsymbol{u}_{j,y}$. 
\end{itemize}
A visual example of the 2D projection acquisition is shown in Fig.~\ref{fig:short}. It is of interest to note that due to the constant intensity assumption of the polyhedron model, $I_{j}(x,y)$ is actually piecewise linear and composed of triangular regions with linear changes of intensity. An example of a 2D projection is shown in Fig.~\ref{fig:example_projection}.

To avoid degeneracy, we assume $\forall k\neq k',\ \boldsymbol{v}_{k}-\boldsymbol{v}_{k'}$ is not colinear with $\boldsymbol{d}_{j}$ to avoid degenerate arrangements of the projected vertices with respect to the direction vector, where two projected vertices coincide with each other.

%The projections may be shifted by an unknown amount. We model this by considering the deviation of the center of the $j$th observation window from the projection of the origin $\boldsymbol{O}_{j}$. The shifts are denoted as $\boldsymbol{s}_{j} = [s_{j,x},s_{j,y}]^{T}\in\mathbb{R}^{2}$ along  $\boldsymbol{u}_{j,x}$ and $\boldsymbol{u}_{j,y}$ directions respectively, as shown in Fig.~\ref{fig:short}.

Moreover, to model the blurring effect of point spread functions and discrete sensors during the measurement stage, we consider the projection $I_{j}(x,y)$ to be filtered with a 2D sampling kernel $\varphi(x,y)$, and the measurements are taken on a uniform grid with step sizes $T_{x}$ and $T_{y}$ along the directions $\boldsymbol{u}_{j,x}$ and $\boldsymbol{u}_{j,y}$ respectively. Hence, we observe the following 2D samples:
\begin{align} \label{eq_sample}
     \nonumber I_{j}[m,n] &=\int_{-\infty}^{\infty}\int_{-\infty}^{\infty}I_{j}(x,y)\varphi\left(\frac{x}{T_{x}}-m,\frac{y}{T_{y}}-n\right)dxdy \\
     &= \biggl\langle I_{j}(x,y), \varphi\left(\frac{x}{T_{x}}-m, \frac{y}{T_{y}} -n\right)\biggr\rangle,
\end{align}
where $m,n\in \mathbb{N}$, and $T_{x},T_{y}\in\mathbb{R}$ are the sampling intervals. Here $\langle\cdot,\cdot\rangle$ denotes inner product. For the rest of this paper, we assume $T_{x}=T_{y}=T$ without loss of generality. 

Given only 2D samples of the projections $I_{j}[m,n],\ j = 1,...,J$, the problem we consider is the simultaneous estimation of the 3D polyhedron $g(\boldsymbol{r})$, the projection direction vectors $\{\boldsymbol{d}_{j}\}_{j=1}^{J}$ and the shifts $\{\boldsymbol{s}_{j}\}_{j=1}^{J}$.

In what follows, we show that the  above problem admits a unique solution up to a 3D orthogonal transformation for polyhedra with $K\geq 4$ vertices, when $J\geq 3$ and the sampling kernel $\varphi(x,y)$ is a polynomial reproducing kernel up to order $2K-4$ along both axes.

\section{Exact estimation using samples on projections from unknown projection angles} \label{section3}

\begin{figure}[t]
\centering
\includegraphics[width=0.35\textwidth]{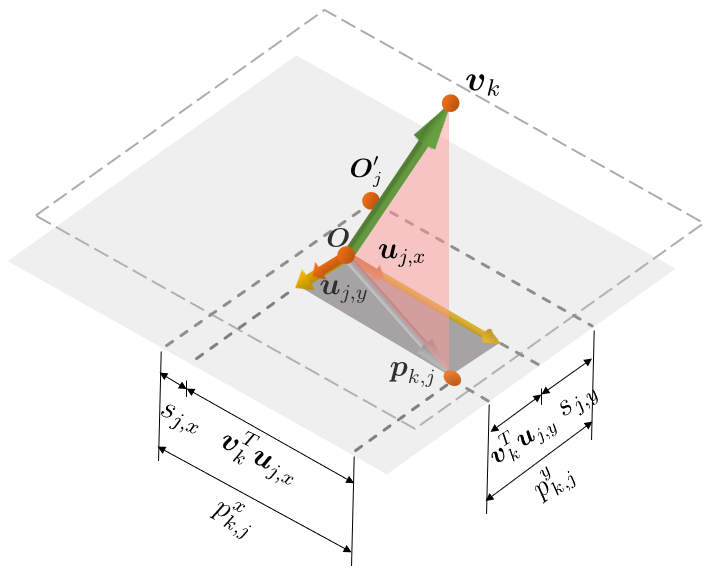}
\caption{An illustration of the geometric relation. The grey plane represents a parallel plane to $\Pi_{j}$ passing through the origin $\boldsymbol{O}$. The transparent plane with dashed line boundaries represents the shifted observation window whose center $\boldsymbol{O}_{j}'$ is shifted by $\boldsymbol{s}_{j}$. The inner product of $\boldsymbol{v}_{k}$ (green vector) with $\boldsymbol{u}_{j,x}$ (orange vector) gives $p_{k,j}^{x}$ up to the shift $s_{j,x}$. One can similarly obtain $p_{k,j}^{y}$.}
\label{fig:geometric_relation}
%\vspace*{-0.5cm}
\end{figure}

Due to convexity, there is a unique way in which the vertices $\boldsymbol{v}_{k}$ can be connected to form the polyhedron $g(\boldsymbol{r})$~\cite{milanfer}. Hence, estimating $g(\boldsymbol{r})$ is equivalent to estimating the locations of its vertices in $\mathbb{R}^{3}$ space.
We denote the projected location of $\boldsymbol{v}_{k}$ on the plane $\Pi_{j}$ as $\boldsymbol{p}_{k,j}=[p_{k,j}^{x},p_{k,j}^{y}]^{T}\in\mathbb{R}^{2}$ (see Fig.~\ref{fig:geometric_relation}).

Moreover, from geometrical properties of the parallel beam projection, $p_{k,j}^{x},\ p_{k,j}^{y}$ can be expressed as the projection of the vertex $\boldsymbol{v}_{k}$ onto the direction vectors $\boldsymbol{u}_{j,x}$ and $\boldsymbol{u}_{j,y}$, then shifted by $s_{j,x}$ and $s_{j,y}$ along the two directions respectively:
 \begin{align} \label{eq5}
     \nonumber\boldsymbol{p}_{k,j} &= [p_{k,j}^{x}, p_{k,j}^{y}]^{T}\\
     &= [\boldsymbol{v}_{k}^{T}\boldsymbol{u}_{j,x}+
s_{j,x}, \boldsymbol{v}_{k}^{T}\boldsymbol{u}_{j,y}+s_{j,y}]^{T}. 
 \end{align}
The above geometric relation is illustrated in Fig.~\ref{fig:geometric_relation}. It links the desired 3D vectors $\boldsymbol{v}_{k},\ \boldsymbol{u}_{j,x}$ and $\boldsymbol{u}_{j,y}$ with 2D parameters $\boldsymbol{p}_{k,j}$, and provides valuable insights on how we can unveil the 3D geometric information from lower dimensional features.

\begin{figure}[t]
\centering
\includegraphics[width=0.48\textwidth]{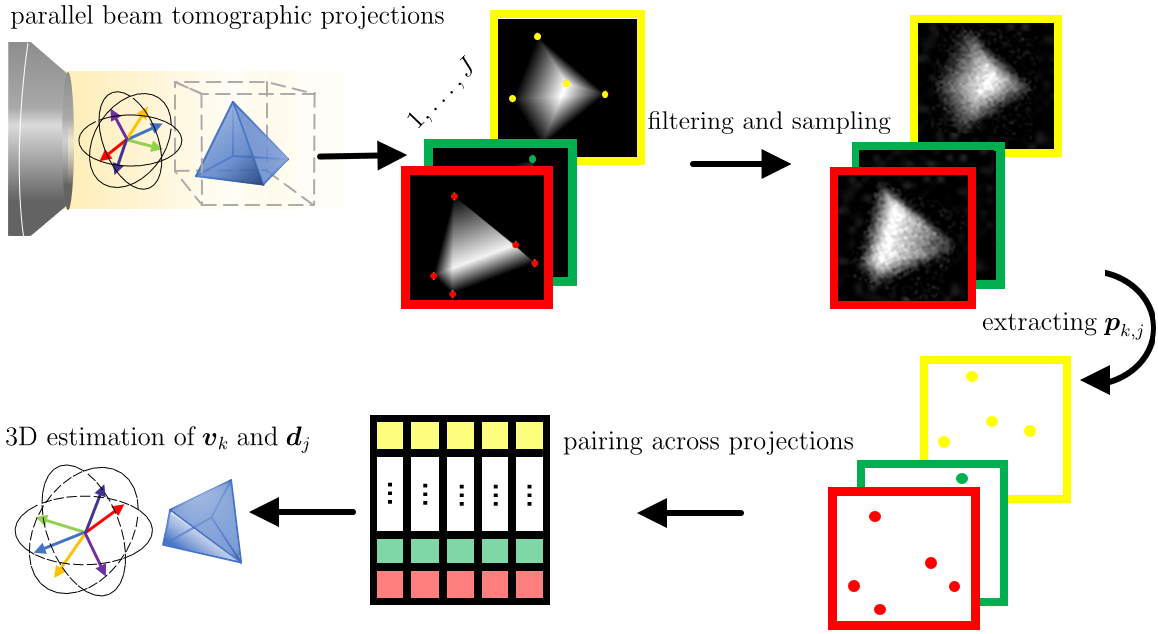}
\caption{An illustration of the overall method. The features $\boldsymbol{p}_{k,j}$ are first extracted from the sampled projections (Sec~\ref{section3sub1}). Then they are paired across different projections (Sec~\ref{section3sub2}). Finally, the 3D polyhedron and projection directions are estimated from the ordered features using an algebraic approach described in Sec~\ref{section3sub3}.}
\label{fig:overall_method}
\end{figure}

In what follows, we provide a constructive solution to the stated reconstruction problem. As a first step, we address the 2D problem of extracting the 2D parameters $\boldsymbol{p}_{k,j}$ from the samples $I_{j}[m,n]$. Then, we show that the desired 3D vectors can be estimated simultaneously by leveraging the geometric relations in Eq.~(\ref{eq5}). An overview of the proposed method is illustrated in Fig.~\ref{fig:overall_method}.

\subsection{Estimation of $\boldsymbol{p}_{k,j}$} \label{section3sub1}

%We start by observing that, under the assumption that $g(\boldsymbol{r})$ is constant within its convex hull, the projection $I_{j}(x,y)$ is piecewise linear. Moreover, since convexity is preserved through projection, $I_{j}(x,y)$ is actually composed of linear polynomial regions $A_{1},A_{2},...,A_{L}$ with convex polygonal boundaries as shown in Fig.~\ref{fig:example_projection}. 
%Geometrically, part of the vertices of the linear polygonal regions $\{A_{l}\}_{l=1}^{L}$ correspond to the projected vertices $\boldsymbol{p}_{k,j}$ as shown in Fig.~\ref{fig:example_projection}. 

We first establish a connection between the projection $I_{j}(x,y)$ and the projected locations $\boldsymbol{p}_{k,j}$ by stating the following lemma:
\begin{lemma}
%Let $h(z)$ be a complex analytic function. For a projection $I_{j}(x,y)$ obtained from Eq.~(\ref{eq3}) with $g(\boldsymbol{r})$ given in Eq.~(\ref{eq1}), there exist complex coefficients $\rho_{1},\rho_{2},...,\rho_{K}\in\mathbb{C}$, such that:
There exist coefficients $\rho_{1},\ \rho_{2},...,\rho_{K}\in\mathbb{C}$, such that for any complex function $h(z)$ analytic within the projection area of $I_{j}(x,y)$ obtained from Eq.~(\ref{eq3}) with $g(\boldsymbol{r})$ given in Eq.~(\ref{eq1}), the following equality holds:
\begin{equation}\label{eq6}
    \iint_{\Pi_{j}}I_{j}(x,y)h'''(z)dxdy = \sum_{k=1}^{K}\rho_{k}h(z_{k,j}),
\end{equation}
where $z_{k,j}=p_{k,j}^{x}+i p_{k,j}^{y}$ corresponds to the locations of $\boldsymbol{p}_{k,j}$ in the complex plane, and $h'''(\cdot)$ is the third order derivative of $h(\cdot)$. 
\label{lemma1_statement}
\end{lemma}
\begin{proof} 
See Appendix~\ref{proof_lemma1}.
\end{proof}
From Lemma~\ref{lemma1_statement}, we understand the importance of finding a proper analytic function $h(z)$ that enables the estimation of $z_{k,j}$ from the \textit{rhs} expression in Eq.~(\ref{eq6}).
%, as well as to approximate exactly the \textit{lhs} integral of Eq.~(\ref{eq6}) using the samples $I_{j}[m,n]$. 
For this purpose, we choose $h(z) = z^{r}$. By doing so, the estimation task is converted to a spectral analysis problem.
%While for the second problem, we leverage the finite rate of innovation sampling theorem and consider $h(z)$ to be reproducible from some basis functions~\cite{pina1,pld1}.
Alternative choices of $h(z)$ will be discussed in Section~\ref{section4}.
By choosing $h(z) = z^{r}$, Eq.~(\ref{eq6}) becomes:
\begin{align} \label{eq7}
    \nonumber&\iint_{\Pi_{j}}I_{j}(x,y)h'''(z)dxdy
    =\iint_{\Pi_{j}}I_{j}(x,y)(z^{r})'''dxdy\\
    \nonumber&=r(r-1)(r-2)\iint_{\Pi_{j}}I_{j}(x,y)(x+iy)^{r-3}dxdy\\
    \nonumber&=r(r-1)(r-2)\tau_{r-3}=\tilde{\tau}_{r}\\
    &=\sum_{k=1}^{K}\rho_{k}z_{k,j}^{r},
\end{align}
where 
\begin{equation*} 
\tau_{r}=\iint_{\Pi_{j}}I_{j}(x,y)(x+iy)^{r}dxdy
\end{equation*}
is the complex moment of $I_{j}(x,y)$, while $\tilde{\tau}_{r}$ is the weighted version of $\tau_{r}$ with weight $r(r-1)(r-2)$. Note that by construction $\tilde{\tau}_{0}=\tilde{\tau}_{1}=\hat{\tau}_{2} = 0$. From Eq.~(\ref{eq7}), it is clear that $\tilde{\tau}_{r}$ are the linear combinations of exponentials of $z_{k,j}$. Thus, we can apply the annihilating filter method (aka Prony's method)~\cite{pina1,prony} to retrieve the $K$ values of $z_{k,j}$ from $2K$ weighted complex moment $\tilde{\tau}_{r},\ r=0,1,...,2K-1$, or equivalently from $\left(2K -3\right)$ complex moments $\tau_{r},\ r = 0,1,...,2K-4$. 

The remaining issue is to obtain the complex moments $\tau_{r}$ from the 2D samples $I_{j}[m,n]$. 
In order to achieve this, we require the sampling kernel $\varphi(x,y)$ in Eq.~(\ref{eq_sample}) to be able to reproduce polynomials up to degree $x^{2K-4}y^{2K-4}$.  In other words, there exist coefficients $c_{m,n}^{\alpha,\beta}$ such that for $0\leq \alpha,\beta\leq 2K-4$:
\begin{equation} \label{eq8}
\sum_{m,n}c_{m,n}^{\alpha,\beta}\varphi\left(\frac{x}{T}-m,\frac{y}{T}-n\right) = x^{\alpha}y^{\beta}.
\end{equation}
Kernels satisfying the above conditions include polynomial splines, and we refer to~\cite{pld1}
for more details.
Leveraging the polynomial reproducing property of the sampling kernel, we can retrieve the geometric moments $\mu_{\alpha,\beta}$ from the 2D samples $I_{j}[m,n]$ as follows~\cite{shukla,me2}:
\begin{align} \label{eq9}
   \nonumber &\sum_{m,n}c_{m,n}^{\alpha,\beta}I_{j}[m,n]\\
   \nonumber&\overset{(b)}{=}\sum_{m,n}c_{m,n}^{\alpha,\beta}\iint_{\Pi_{j}}I_{j}(x,y)\varphi\left(\frac{x}{T}-m,\frac{y}{T}-n\right)dxdy\\
  \nonumber  &=\iint_{\Pi_{j}}I_{j}(x,y)\sum_{m,n}c_{m,n}^{\alpha,\beta}\varphi\left(\frac{x}{T}-m,\frac{y}{T}-n\right)dxdy\\
  \nonumber  &\overset{(c)}{=}\iint_{\Pi_{j}}I_{j}(x,y)x^{\alpha}y^{\beta}dxdy\\
    &=\mu_{\alpha,\beta},
\end{align}
where $(b)$ and $(c)$ follows from Eq.~(\ref{eq_sample}) and Eq.~(\ref{eq8}) respectively. Finally, the geometric moments $\mu_{\alpha,\beta}$ relate to the complex moments $\tau_{r}$ through binomial expansion:
\begin{equation}\label{moment_definition}
    \tau_{r} = \sum_{\beta=0}^{r} \binom{r}{\beta}i^{\beta}\mu_{\alpha,\beta},\  \alpha+\beta = r.
\end{equation}

Based on the above derivations, we now have an algorithm to estimate $\boldsymbol{p}_{k,j}$ from the sampled projection $I_{j}[m,n]$:
\begin{enumerate}
    \item Compute the geometric moments $\mu_{\alpha,\beta}$ from the samples $I_{j}[m,n]$ using Eq.~(\ref{eq9}).
    \item Compute the complex moment $\tau_r$ using Eq.~(\ref{moment_definition}).
    \item Using the relationship in Eq.~(\ref{eq7}): 
    $$r(r-1)(r-2)\tau_{r-3}= \tilde{\tau}_{r}= \sum_{k=1}^{K}\rho_{k}z_{k,j}^{r},$$ apply Prony's method to
    $\tilde{\tau}_{r}$ to retrieve the locations $\boldsymbol{p}_{k,j}$.
\end{enumerate}

Once the 2D parameters $\{\boldsymbol{p}_{k,j}\}_{k=1}^{K}$ are retrieved, we can estimate the unknown shifts $\boldsymbol{s}_{j}$ on each projected plane by summing the retrieved parameters for all vertices as follows:
\begin{align}\label{eq10}
    \nonumber \sum_{k=1}^{K}p_{k,j}^{x}&\overset{(d)}{=} \left(\sum_{k=1}^{K}\boldsymbol{v}_{k}^{T}\right)\boldsymbol{u}_{j,x}+K s_{j,x}\overset{(e)}{=}K s_{j,x}\\
     \sum_{k=1}^{K}p_{k,j}^{y}&\overset{(d)}{=} \left(\sum_{k=1}^{K}\boldsymbol{v}_{k}^{T}\right)\boldsymbol{u}_{j,y}+K s_{j,y}\overset{(e)}{=}K s_{j,y},
\end{align}
where $(d)$ follows from Eq.~(\ref{eq5}) and $(e)$ follows from the assumption that the geometric center of the vertices is at the origin.
We denote the shift corrected parameters as $\boldsymbol{\hat{p}}_{k,j}$ with $\boldsymbol{\hat{p}}_{k,j}=\boldsymbol{p}_{k,j}-\boldsymbol{s}_{j}$.

\subsection{Pairing $\boldsymbol{\hat{p}}_{k,j}$ with $\boldsymbol{v}_{k}$} 
\label{section3sub2}

Given the values of $\boldsymbol{\hat{p}}_{k,j}$, we need to identify whether two parameters $\boldsymbol{\hat{p}}_{k,j}$ and $\boldsymbol{\hat{p}}_{k',l}$ retrieved from two arbitrary projections $I_{j}[m,n]$ and $I_{l}[m,n]$ correspond to the same vertex $\boldsymbol{v}_{k}$. 

We now show that this can be done in a pairwise manner. Given two sets of unpaired parameters $\{\boldsymbol{\hat{p}}_{k,j}\}_{k=1}^{K}$ and $\{\boldsymbol{\hat{p}}_{k',l}\}_{k'=1}^{K}$ related to arbitrary projections $I_{j}(x,y)$ and $I_{l}(x,y)$, we build the following matrix $\Upsilon\in\mathbb{R}^{K\times 4}$:
\begin{equation}\label{eq11}
\Upsilon = 
    \begin{bmatrix}
        \boldsymbol{\hat{p}}_{1,j}^{T}&\boldsymbol{\hat{p}}_{1,l}^{T}\\
         \boldsymbol{\hat{p}}_{2,j}^{T}&\boldsymbol{\hat{p}}_{2,l}^{T}\\
        \vdots&\vdots\\
         \boldsymbol{\hat{p}}_{K,j}^{T}&\boldsymbol{\hat{p}}_{K,l}^{T}\\
    \end{bmatrix}
\end{equation}
and we make the following observation:
\begin{lemma} \label{lemma_pairing}
    The matrix $\Upsilon\in\mathbb{R}^{K\times 4}$ with $K\geq 4$ is rank deficient only when the entries along the same row $k$ correspond to the same vertex $\boldsymbol{v}_{k}$.
\end{lemma}
\begin{proof}
When the correct pairing is reached, the matrix $\Upsilon$ can be factorized as follows:
\begin{equation*}
    \Upsilon \overset{(f)}{=} 
    \underbrace{
    \begin{bmatrix}
        \boldsymbol{v}_{1}^{T}\\
         \boldsymbol{v}_{2}^{T}\\
         \vdots\\
         \boldsymbol{v}_{K}^{T}
    \end{bmatrix}}_{K\times 3}
    \underbrace{
    \begin{bmatrix}
        \boldsymbol{u}_{j,x}&\boldsymbol{u}_{j,y}& \boldsymbol{u}_{l,x}&\boldsymbol{u}_{l,y}
    \end{bmatrix}
    }_{3\times 4}=\boldsymbol{SC},
\end{equation*}
where $(f)$ follows from Eq.~(\ref{eq5}).
By construction, the matrices $\boldsymbol{S}$ and $\boldsymbol{C}$ are of rank 3, which means the rank of $\Upsilon$ is at most $3$. On the contrary, when the correct pairing is not reached, the factorization cannot be performed, which results in a full rank matrix $\Upsilon$, that is rank($\Upsilon$) = 4. 
\end{proof}
Lemma~\ref{lemma_pairing} provides a way to pair the retrieved parameters $\{\boldsymbol{\hat{p}}_{k,j}\}_{k=1}^{K}$ and $\{\boldsymbol{\hat{p}}_{k',l}\}_{k'=1}^{K}$ from two projections. Specifically, when $K\geq 4$, we can perform the pairing between two projections by permuting the order of one set of parameters, e.g. permuting $\{\boldsymbol{\hat{p}}_{k',l}\}_{k'=1}^{K}$. For each permutation, we can construct a matrix $\Upsilon$ using Eq.~(\ref{eq11}). There are in total $K!$ possible permutations, and we choose the permutation that corresponds to the matrix $\Upsilon$ with a deficient rank. 

%\vspace*{-0.5cm}

\subsection{Estimation of $\boldsymbol{d}_{j}$ and $\boldsymbol{v}_{k}$} \label{section3sub3}

We are now in the position to state the sampling theorem for an exact estimation of the 3D polyhedron in Eq.~(\ref{eq1}) from samples of its 2D tomographic projections at unknown angles in Eq.~(\ref{eq_sample}):
\begin{theorem}
\label{theorem1}
Given uniform samples $I_{j}[m,n],\ j=1,...,J$, of $J\geq 3$     projections at unknown projection angles, the projection angles and any 3D bilevel convex polyhedron with $K\geq 4$ vertices can be exactly reconstructed up to a 3D orthogonal transformation when the sampling kernel $\varphi(x,y)$ is a polynomial reproducing kernel up to order $2K-4$ along both axes. 
%or an exponential reproducing kernel up to order $2K-1$ along both axes.
\end{theorem}

\begin{proof}
Given that the sampling kernel $\varphi(x,y)$ can reproduce 2D polynomials up to degree $2K-4$ along both axes, by considering Lemma~\ref{lemma1_statement} with a specific choice of the analytic function $h(z) = z^{r}$, the 2D parameters $\boldsymbol{\hat{p}}_{k,j}$ can be retrieved exactly for every 2D projection as in Section~\ref{section3sub1}. Then, given $\boldsymbol{\hat{p}}_{k,j}$, we perform the pairing by building the matrix $\Upsilon\in\mathbb{R}^{K\times 4}$ in Eq.~(\ref{eq11}) and applying the rank criterion in Lemma~\ref{lemma_pairing}.

After pairing the parameters $\boldsymbol{\hat{p}}_{k,j}$, we now show that it is possible to estimate exactly the projection directions $\boldsymbol{d}_{j}$ and the locations of the vertices $\boldsymbol{v}_{k}$ up to an orthogonal transformation, using a factorization approach similar to the one proposed in \cite{tomasi}. In what follows, we provide a detailed explanation of the steps involved in the approach. 

Using only the first element of the paired parameters $\{\hat{p}_{k,j}^{x}\}_{k=1,j=1}^{K,J}$, we build the following matrix $\boldsymbol{\Omega}_{x}\in\mathbb{R}^{K\times J}$:
\begin{align} \label{eq12}
   \nonumber \boldsymbol{\Omega}_{x} &= 
    \begin{bmatrix}
        \hat{p}_{1,1}^{x}&\hat{p}_{1,2}^{x}&\hdots&\hat{p}_{1,J}^{x}\\
        \hat{p}_{2,1}^{x}&\hat{p}_{2,2}^{x}&\hdots&\hat{p}_{2,J}^{x}\\
        \vdots&\vdots&\ddots&\vdots\\
        \hat{p}_{K,1}^{x}&\hat{p}_{K,2}^{x}&\hdots&\hat{p}_{K,J}^{x}
    \end{bmatrix}_{K\times J}\\
    &=\underbrace{
    \begin{bmatrix}
        \boldsymbol{v}_{1}^{T}\\
         \boldsymbol{v}_{2}^{T}\\
         \vdots\\
         \boldsymbol{v}_{K}^{T}
    \end{bmatrix}}_{K\times 3}
    \underbrace{
    \begin{bmatrix}
    \boldsymbol{u}_{1,x}&\boldsymbol{u}_{2,x}&\hdots&\boldsymbol{u}_{J,x}
    \end{bmatrix}}_{3\times J}=\boldsymbol{V}\boldsymbol{U}_{x}.
\end{align}
By construction, the matrices $\boldsymbol{V}$ and $\boldsymbol{U}_{x}$ are of dimensions $K\times3$ and $3\times J$ respectively. Therefore, the matrix $\boldsymbol{\Omega}_{x}$ has rank $\leq 3$.
Similarly, we build another rank deficient matrix $\boldsymbol{\Omega}_{y}$ using the second element in the paired parameters $\{\hat{p}_{k,j}^{y}\}_{k=1,j=1}^{K,J}$:
\begin{equation}\label{eq13}
    \boldsymbol{\Omega}_{y}=
    \begin{bmatrix}
        \boldsymbol{v}_{1}^{T}\\
         \boldsymbol{v}_{2}^{T}\\
         \vdots\\
         \boldsymbol{v}_{K}^{T}
    \end{bmatrix}
    \begin{bmatrix}
    \boldsymbol{u}_{1,y}&\boldsymbol{u}_{2,y}&\hdots&\boldsymbol{u}_{J,y}
    \end{bmatrix}=\boldsymbol{V}\boldsymbol{U}_{y}.
\end{equation}
Similar to $\boldsymbol{\Omega}_{x}$, $\boldsymbol{\Omega}_{y}$ is also of rank $\leq 3$. 

We first concatenate the two matrices: $\boldsymbol{\Omega}= \begin{bmatrix}
\boldsymbol{\Omega}_{x}\ |\ \boldsymbol{\Omega}_{y}
\end{bmatrix}$, such that $\boldsymbol{\Omega}\in\mathbb{R}^{K\times 2J}$. It follows that
\begin{equation} \label{definition_omega}
    \boldsymbol{\Omega} = \boldsymbol{V}
    \begin{bmatrix}
        \boldsymbol{U}_{x}\ |\ \boldsymbol{U}_{y}
    \end{bmatrix},\text{ and rank}(\boldsymbol{\Omega})\leq 3, 
\end{equation}
where $\begin{bmatrix}
    \boldsymbol{U}_{x}\ |\ \boldsymbol{U}_{y}
\end{bmatrix} \in\mathbb{R}^{3\times 2J}$. Then we perform singular value decomposition on the matrix: $\boldsymbol{\Omega} = \mathcal{US V}^{T}$. The initial estimation of the direction vector matrices $\boldsymbol{U}_{x}, \boldsymbol{U}_{y}$ can be obtained as:
\begin{equation*}
    \begin{bmatrix}
        \boldsymbol{\hat{U}}_{x}\ |\ \boldsymbol{\hat{U}}_{y}
    \end{bmatrix} = \mathcal{S V}^{T}=
    \begin{bmatrix}
    \boldsymbol{\hat{u}}_{1,x}\ \hdots\ \boldsymbol{\hat{u}}_{J,x}\ |\ \boldsymbol{\hat{u}}_{1,y}\ \hdots\ \boldsymbol{\hat{u}}_{J,y}
    \end{bmatrix}.
\end{equation*}
Clearly, the initial estimation $\begin{bmatrix}
    \boldsymbol{\hat{U}}_{x}\ | \ \boldsymbol{\hat{U}}_{y}
\end{bmatrix}$ relates to the true direction vectors $\begin{bmatrix}
    \boldsymbol{U}_{x}\ |\ \boldsymbol{U}_{y}
\end{bmatrix}$ by an unknown linear transformation, which we denote with $\boldsymbol{Q}\in\mathbb{R}^{3\times 3}$: 
\begin{equation}\label{eq14}
    \boldsymbol{Q}\begin{bmatrix}
        \boldsymbol{\hat{U}}_{x}\ |\ \boldsymbol{\hat{U}}_{y}
    \end{bmatrix} = \begin{bmatrix}
        \boldsymbol{U}_{x}\ |\ \boldsymbol{U}_{y}
    \end{bmatrix}.
\end{equation}
We identify the linear transformation $\boldsymbol{Q}$ by considering the fact that the matrix $\begin{bmatrix}
\boldsymbol{U}_{x}\ |\ \boldsymbol{U}_{y}
\end{bmatrix}$ has unit norm columns: $\boldsymbol{u}_{j,x}^{T}\boldsymbol{u}_{j,x}=1$ and $\boldsymbol{u}_{j,y}^{T}\boldsymbol{u}_{j,y}=1$. We replace $\boldsymbol{u}_{j,x}, \boldsymbol{u}_{j,y}$ with $\boldsymbol{Q}\boldsymbol{\hat{u}}_{j,x}$ and $\boldsymbol{Q}\boldsymbol{\hat{u}}_{j,y}$, and this yields:
\begin{equation}\label{eq15}
\boldsymbol{\hat{u}}_{j,x}^{T}\underbrace{\boldsymbol{Q}^{T}\boldsymbol{Q}}_{\boldsymbol{M}}\boldsymbol{\hat{u}}_{j,x}=1\text{ and } \boldsymbol{\hat{u}}_{j,y}^{T}\underbrace{\boldsymbol{Q}^{T}\boldsymbol{Q}}_{\boldsymbol{M}}\boldsymbol{\hat{u}}_{j,y}=1,
\end{equation}
where $\boldsymbol{M} = \boldsymbol{Q}^{T}\boldsymbol{Q}$ is a symmetric matrix of dimension $3\times 3$. Hence, it
\begin{comment}
\begin{equation*}
    \boldsymbol{M} = 
    \begin{bmatrix}
    a&b&c\\
    b&a&d\\
    c&d&a
    \end{bmatrix},
\end{equation*}
\end{comment}
has six unknown values. Under the assumption that the vectors in $\{\boldsymbol{d}_{j}\}_{j=1}^{J}$, $\{\boldsymbol{u}_{j,x}\}_{j=1}^{J}$ and $\{\boldsymbol{u}_{j,y}\}_{j=1}^{J}$ are pairwise distinct, given $J\geq 3$ projections, we can solve uniquely for $\boldsymbol{M}$ using the system of $2J$ linear equations in Eq.~(\ref{eq15}). The linear transformation $\boldsymbol{Q}$ is then given by $\boldsymbol{B}(\boldsymbol{M})^{\frac{1}{2}}$, where $\boldsymbol{B}$ is an arbitrary orthogonal matrix of dimension $3\times 3$, since $\boldsymbol{B}^{T}\boldsymbol{B} = \boldsymbol{I}$. Then, the final estimation of $\boldsymbol{\hat{u}}_{j,x}$ and $\boldsymbol{\hat{u}}_{j,y}$ can be obtained by applying the linear transformation $\boldsymbol{Q}$ using Eq.~(\ref{eq14}).

Consequently, the vertices $\boldsymbol{\hat{v}}_{k}$ can be retrieved by solving a linear system of equations using Eq.~(\ref{eq12}) and (\ref{eq13}). Finally, given $\boldsymbol{\hat{u}}_{j,x}$ and $\boldsymbol{\hat{u}}_{j,y}$, the direction vectors $\boldsymbol{\hat{d}}_{j}$ can be retrieved using the relation  in Eq.~(\ref{eq2}).  

Due to the existence of an arbitrary orthogonal matrix $\boldsymbol{B}$ in the final solution, the estimation of the locations of $\boldsymbol{\hat{v}}_{k}$ and the projection directions $\boldsymbol{\hat{d}}_{j}$ will be up to an orthogonal transformation from the true values.
\end{proof}
We summarize the complete method in Algorithm 1.

\textit{Remark 1.} The above reconstruction method can be applied to the estimation of 3D point sources from projections at unknown angles. The locations and amplitudes of the projected point sources can be retrieved from the 2D samples using techniques in the area of finite rate of innovation~\cite{shukla, hajat, hanjie1}. Given the assumptions as in Theorem~\ref{theorem1}, the projection directions and the locations of point sources can be estimated up to an orthogonal transformation using the proposed method.

\textit{Remark 2.} The perfect reconstruction of 2D signals from their 1D tomographic projections can also be achieved. For 2D point sources and bilevel convex polygon models, given that there are $K\geq 3$ non-colinear vertices and the projections are taken at $J\geq 3$ different angles, the estimation can be achieved up to a 2D orthogonal transformation. 

\textit{Remark 3.} If three direction vectors in $\boldsymbol{U}_{x}$ or $\boldsymbol{U}_{y}$ are known, we can determine the orthogonal matrix $\boldsymbol{B}$ and estimate the locations of the vertices and the projection angles exactly.

\begin{algorithm}
 \caption{Algorithm for reconstructing the 3D polyhedron}
 \begin{algorithmic}[1]
 \renewcommand{\algorithmicrequire}{\textbf{Input:}}
 \renewcommand{\algorithmicensure}{\textbf{Output:}}
 \REQUIRE 2D samples $\{I_{j}[m,n]\}_{j=1}^{J}$, $J\geq 3$ and $K\geq 4$
 \ENSURE 3D vertices $\{\boldsymbol{\hat{v}}_{k}\}_{k=1}^{K}$,  projection directions $\{\boldsymbol{\hat{d}}_{j}\}_{j=1}^{J}$ and 2D shifts $\boldsymbol{\hat{s}}_{j}$
  \STATE For every projection, compute geometric moments $\mu_{\alpha,\beta}$ using Eq.~(\ref{eq9}).
  \STATE Compute weighted complex moment $\tilde{\tau}_{r}$ according to Eq.~(\ref{moment_definition}) and Eq.~(\ref{eq7}).
  \STATE Apply annihilating filter method to $\tilde{\tau}_{r}$ to retrieve 2D parameters $\boldsymbol{p}_{k,j}$.
  \STATE Estimate in-plane shifts $\boldsymbol{s}_{j}$ according to Eq.~(\ref{eq10}).
  \STATE Pair corrected parameters $\boldsymbol{\hat{p}}_{k,j}$ using the rank criterion in Lemma~\ref{lemma_pairing}.
  \STATE Estimate the direction vectors $\{\boldsymbol{\hat{u}}_{j,x}\}_{j=1}^{J}$ and $\{\boldsymbol{\hat{u}}_{j,y}\}_{j=1}^{J}$ by performing SVD on $\boldsymbol{\Omega}$ in Eq.~(\ref{definition_omega}).
  \STATE Estimate the linear transform $\boldsymbol{Q}$ by solving the linear system of equations in Eq.~(\ref{eq15}).
  \STATE Apply the transform $\boldsymbol{Q}$ to the initial estimation of $\{\boldsymbol{\hat{u}}_{j,x}\}_{j=1}^{J}$ and $\{\boldsymbol{\hat{u}}_{j,y}\}_{j=1}^{J}$ to get the final estimation as in Eq.~(\ref{eq14}).
  \STATE Estimate 3D vertices $\{\boldsymbol{\hat{v}}_{k}\}_{k=1}^{K}$ by solving the linear system of equations in Eq.~(\ref{eq13}) and Eq.~(\ref{eq12}).
  \STATE Compute the projection direction $\{\boldsymbol{\hat{d}}_{j}\}_{j=1}^{J}$ according to Eq.~(\ref{eq2}).
 \end{algorithmic} 
 \end{algorithm}

\section{Noisy setting} \label{section4}

We now assume the 2D samples $I_{j}[m,n]$ are corrupted by additive, white noise. Therefore, we measure $\tilde{I}_{j}[m,n] = I_{j}[m,n] + \epsilon_{m,n}$, where $\epsilon_{m,n}$ is the additive noise. In this section, we introduce alternative strategies to mitigate the effect of noise, including robust estimation of the 2D parameters $\boldsymbol{p}_{k,j}$ and the 3D vectors $\boldsymbol{v}_{k}$ and $\boldsymbol{d}_{j}$.

\subsection{2D Parameter Estimation through Analytic Approximation} \label{section4sub1}
We developed Eq.~(\ref{eq6}) in Section~\ref{section3sub1} with the analytic functions $h(z)$ to be complex polynomials.

To improve stability, we now relax the constraint of an exact reproduction of polynomials and opt for a more localized family of analytic functions: $h_{a_{w}}(z) = \frac{1}{z-a_{w}}$, where $a_{w}\in\mathbb{C}$, and $a_{w} = a_{0}e^{i2\pi\frac{w}{W}},\ w = 0,1,...,W-1$. Furthermore, $a_{0}\in\mathbb{R}$ is chosen to satisfy $|a_{0}|> R$ so that the function $h_{a_{w}}(z)$ is analytic in the projection area. By considering Lemma~\ref{lemma1_statement}, we substitute $h_{a_{w}} (z)$ in Eq.~(\ref{eq6}) and get the following equation:
\begin{align*} 
   & \nonumber\iint_{\Pi_{j}}I_{j}(x,y)h_{a_{w}}'''(z)dxdy\\
     &=-6\iint_{\Pi_{j}}I_{j}(x,y)\frac{1}{\left(z-a_{w} \right)^4}dxdy\\
      &= \nonumber \sum_{k=1}^{K}\frac{\rho_{k}}{z_{k,j}-a_{w}}.
\end{align*}
In order to approximate the integral in the above equation, we find coefficients $c_{m,n}^{w}$ such that a linear combination of the sampling kernel $\varphi(x,y)$ and its uniform shifts gives the best approximation of the analytic function $\frac{1}{\left( z-a_{w}\right)^4}$:
\begin{equation}\label{approx_rep}
    \sum_{m,n}c_{m,n}^{w}\varphi\left(\frac{x}{T}-m,\frac{y}{T}-n\right)\cong \frac{1}{\left( z-a_{w}\right)^4}.
\end{equation}
In order to find the optimal coefficients $c_{m,n}^{w}$, we apply the \textit{least squares approximation method} introduced in~\cite{tony}. Therefore, we have the following:
\begin{align}\label{approx_sig_mom}
\nonumber &-6\sum_{m,n} c_{m,n}^{w}I_{j}[m,n]\\
\nonumber &= -6\iint_{\Pi_{j}}I_{j}(x,y)\sum_{m,n}c_{m,n}^{w}\varphi\left(\frac{x}{T}-m,\frac{y}{T}-n\right)dxdy\\
\nonumber &\cong -6\iint_{\Pi_{j}}I_{j}(x,y)\frac{1}{(z-a_{w})^{4}}dxdy \\
 &= \sum_{k=1}^{K}\frac{\rho_{k}}{z_{k,j}-a_{w}}={\eta}_{w}.
\end{align}

Given $\{\eta_{w}\}_{w=0}^{W-1}$, an annihilating filter based solution can be applied to retrieve the positions of the projected vertices $z_{k,j}$ when $W>2K$~\cite{analytic} (please refer to Appendix~\ref{annihilating_filter} for more details). Compared to the choice of analytic functions in Section~\ref{section3sub1}, the advantage of choosing $h_{a_{w}}(z) = \frac{1}{z-a_{w}}$ is that, when the number of the vertices $K$ increases, we only need to place more poles on the circle instead of increasing the polynomial orders as in Eq.~(\ref{eq7}). 
Moreover, the specific form of the chosen analytic function facilitates a robust estimation algorithm. Since $h_{a_{\omega}}(z)$ is a rational function, $\eta_{\omega}$ can be expressed as the ratio of two polynomials:
\begin{equation}\label{eq18}
 \eta_{\omega}=\frac{\sum_{k=0}^{K-1}\lambda_{k}a_{\omega}^{k}}{\prod_{k=1}^{K}(a_{\omega}-z_{k,j})} =\frac{P_{K-1}\left( a_{\omega}\right)}{Q_{K}\left( a_{\omega}\right)}, 
\end{equation}
where $\lambda_{k}\in\mathbb{C}$. The denominator polynomial $Q_{K}(a_{\omega})$ is the annihilating filter whose roots are the positions of the projected vertices $z_{k,j}$. Exploiting the ratio structure, a model-fitting approach can be applied to obtain a more robust estimation of $z_{k,j}$:
\begin{equation}\label{eq19}
    \min_{Q_{K}, P_{K-1}}\sum_{\omega=0}^{W-1}\left|\eta_{\omega}-\frac{P_{K-1}(a_{\omega})}{Q_{K}(a_{\omega})}\right|^2.
\end{equation}
We note that the above minimization problem is non-linear. However, if the poles are placed uniformly on the circle: $a_{\omega} = a_{0}e^{i2\pi\frac{\omega}{W}}$, then we can apply the scheme proposed in~\cite{gilliam1}. Specifically, the minimization scheme estimates the solution of (\ref{eq19}) in a linear manner by considering an iterative quadratic minimization as follows:
\begin{equation*}
     \min_{Q_{K}^{(i)}, P_{K-1}^{(i)}}\sum_{w=0}^{\omega_{\text{max}}-1}\left|\frac{Q_{K}^{(i)}(a_{\omega})\eta_{\omega}-P_{K-1}^{(i)}(a_{\omega})}{Q_{K}^{(i-1)}(a_{\omega})}\right|^2.
\end{equation*}
Each iteration yields a candidate for $Q_{K}^{(i)},i = 1,...,iter_{max}$, and the final solution of $Q_{K}$ is chosen such that the mean square error (MSE) between the resynthesized values of $\eta_{\omega}$ and the measured values is the smallest. Finally, the projected vertices are retrieved as the roots of $Q_{K}$. Please refer to~\cite{gilliam1} for more details.

\begin{figure*}[!h]

\subfigure[]{
\centering
\begin{minipage}[t]{0.3\linewidth}
\centering
\includegraphics[width=2in]{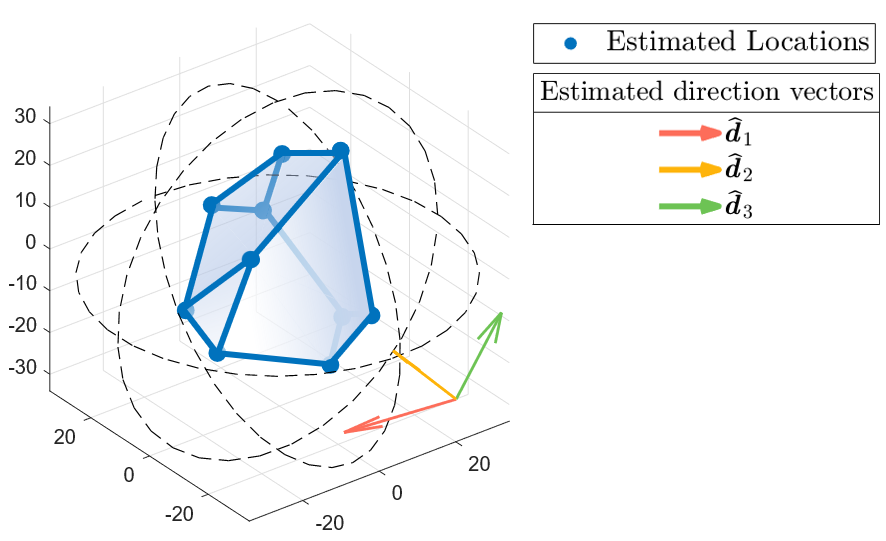}
\end{minipage}%
}%
\subfigure[]{
\centering
\begin{minipage}[t]{0.3\linewidth}
\centering
\includegraphics[width=1.9in]{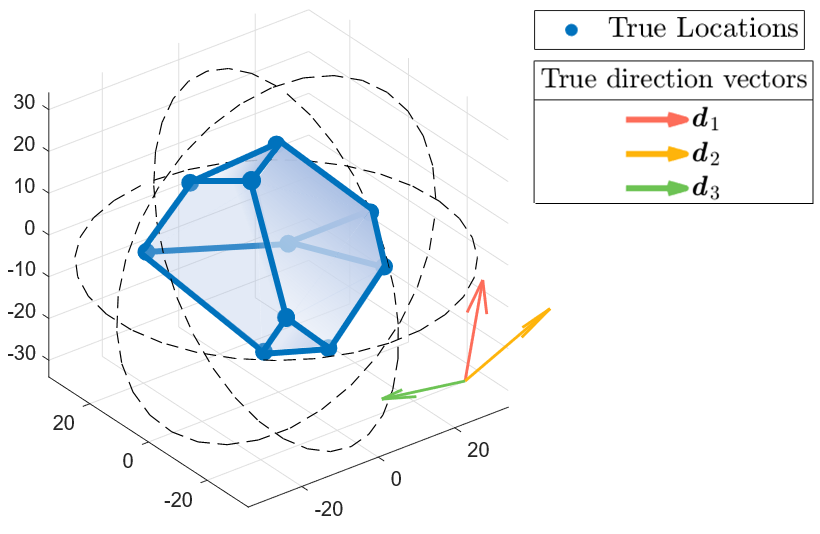}
\end{minipage}%
}%
\ \ 
\subfigure[]{
\begin{minipage}[t]{0.3\linewidth}
\centering
\includegraphics[width=2in]{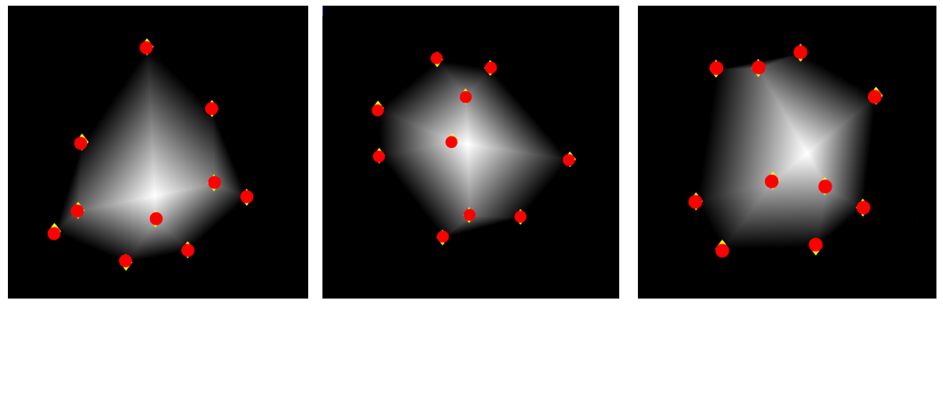}
\end{minipage}%
}%
\centering
\caption{(a) The estimation result using our proposed method is up to an orthogonal transformation compared to the true values in (b). The $3$ projections are shown in (c), while the estimation of the 2D locations of the projected vertices is shown with red dots which overlaps with the true vertices (yellow diamonds). }
\label{fig:perfect}
%\vspace*{-0.4cm}
\end{figure*}

\subsection{Robust estimation of 3D parameters}
\label{section4sub2}
In Section \ref{section3sub3}, once the direction vectors $\boldsymbol{\hat{U}}_{x}$ are retrieved, the estimation of the 3D vertices $\boldsymbol{\hat{V}}$ and the other direction vectors $\boldsymbol{\hat{U}}_{y}$ is based on solving a linear system of equations as in Eq.~(\ref{eq12}) and (\ref{eq13}). In noiseless cases, the solution is accurate. However, in the presence of noise, this may not be the case. Hence, once the initial estimation of $\boldsymbol{\hat{V}}$ and $\boldsymbol{\hat{U}}_{x}$ is obtained using the method in Section \ref{section3sub3}, we further refine it as follows.

For every projection $j$, we first fix $\boldsymbol{\hat{V}}$ and $\boldsymbol{\hat{U}}_{x}$ and formulate the following convex optimization problem to estimate $\boldsymbol{\hat{u}}_{j,y}$:
\begin{align}\label{min1}
    \nonumber \min_{\boldsymbol{\hat{u}}_{j,y}}  & \norm{\boldsymbol{\hat{V}}\boldsymbol{\hat{u}}_{j,y}-\boldsymbol{\hat{p}}_{j}^{y}}^{2},\\
    \nonumber \text{subject to } &\boldsymbol{\hat{u}}_{j,x}^{T}\boldsymbol{\hat{u}}_{j,y} = 0,\\
    &\|\boldsymbol{\hat{u}}_{j,y}\| \leq 1,
\end{align}
where $\boldsymbol{\hat{p}}_{j}^{y}=\left[\hat{p}_{1,j}^{y},\hat{p}_{2,j}^{y},...,\hat{p}_{K,j}^{y}\right]^{T}$. The equality constraint is used to enforce the geometric consistency of the problem and the inequality constraint is a convex relaxation of the norm constraint. A feasible solution for the above minimization problem can be obtained using  disciplined convex programming\footnote{We have used the MATLAB CVX solver (http://cvxr.com/cvx/) which implements the algorithm described in~\cite{convex}.}\cite{convex}. Note that we solve the minimization in (\ref{min1}) for every projection $j$ to estimate all the direction vectors in $\boldsymbol{\hat{U}}_{y}$. 

Next, given $\boldsymbol{\hat{U}}_{x}$ and $\boldsymbol{\hat{U}}_{y}$, we refine the estimation for $\boldsymbol{\hat{V}}$ by solving the following minimization problem:
\begin{equation}\label{min_s}
    \min_{\boldsymbol{\hat{V}}} \norm{\boldsymbol{\hat{V}}\boldsymbol{\hat{U}}_{{x}}-\boldsymbol{\hat{p}}^{x}}^{2}+\norm{\boldsymbol{\hat{V}}\boldsymbol{\hat{U}}_{{y}}-\boldsymbol{\hat{p}}^{y}}^{2},
\end{equation}
where $\boldsymbol{\hat{\boldsymbol{p}}}^{x} = [\boldsymbol{\hat{p}}_{1}^{x},\boldsymbol{\hat{p}}_{2}^{x},...,\boldsymbol{\hat{p}}_{J}^{x}]$ and $\boldsymbol{\hat{p}}^{y} = [\boldsymbol{\hat{p}}_{1}^{y},\boldsymbol{\hat{p}}_{2}^{y},...,\boldsymbol{\hat{p}}_{J}^{y}]$. The above minimization has a closed form solution. Finally, we update the other direction vectors $\boldsymbol{\hat{U}}_{x}$ based on the refined $\boldsymbol{\hat{V}}$ and $\boldsymbol{\hat{U}}_{y}$ in a similar fashion as (\ref{min1}). The above updating process can be repeated to obtain a more precise reconstruction. Normally, a few such iterations lead to a reliable estimation. 

\section{Numerical Simulations and Results}\label{section5}
In this section, we validate the performance of the proposed algorithm, and show that it is able to effectively estimate the 3D polyhedron and projection angles from a small number of sampled and noisy projections. Moreover, we show that aspects of the proposed method can be applied to the geometrical calibration of X-ray scanning systems using real CT data.

\subsection{Evaluation Metrics}
Given the rotational invariance of our problem, in all experiments, once we estimated the 3D vertices $\{\boldsymbol{\hat{v}}_{k}\}_{k=1}^{K}$, we find the orthogonal transformation matrix $\boldsymbol{Q}$ between the estimated 3D locations and the true locations $\{\boldsymbol{v}_{k}\}_{k=1}^{K}$ by considering an orthogonal Procrustes problem~\cite{procrustes}. Then, we quantify the error between the estimated locations and the true locations as:
\begin{equation*}
    E_{\text{vertex}} = \frac{1}{K}\sum_{k=1}^{K} \frac{\|\boldsymbol{Q}{\boldsymbol{\hat{v}}_{k}}-\boldsymbol{v}_{k}\|_{2}}{\|\boldsymbol{v}_{k}\|_{2}}.
\end{equation*}
Using the same orthogonal matrix $\boldsymbol{Q}$, we define the estimation error for direction vectors as follows:

\begin{equation*}
    E_{\text{direction}} = \frac{1}{J}\sum_{j=1}^{J} \left(1-\big\langle \boldsymbol{Q}{\boldsymbol{\hat{d}}_{j}},\ \boldsymbol{d}_{j}\big\rangle\right).
\end{equation*}

\subsection{Noiseless Setting}
Fig.~\ref{fig:perfect} (a) shows the estimation result for a convex bilevel polyhedron with $K = 10$ vertices from 2D samples of $J=3$ tomographic projections taken at unknown directions. The size of the rectangular observation window is assumed to be $(2.5R,\ 2.5R)$, $R=32$, and the unknown shifts $\boldsymbol{s}_{j}$ are randomly drawn from the Uniform distribution $\mathcal{U}(-0.1R,\ 0.1R)$. The 2D projections are of size $1024\times1024$ pixels. Their corresponding 2D samples are generated from filtering the noise-free 2D tomographic projections in Fig.~\ref{fig:perfect} (c) with a 2D B-spline kernel of order $16$. As expected, when compared to the true polyhedron and direction vectors in Fig.~\ref{fig:perfect}~(b), the reconstruction in Fig.~\ref{fig:perfect}~(a) is up to an orthogonal transformation. After applying the orthogonal matrix $\boldsymbol{Q}$, the reconstruction errors in the locations and directions are $E_{\text{vertex}} = 8.10\times 10^{-3}$ and $E_{\text{direction}}=5.61\times 10^{-3}$). 

\begin{figure}[!t]
\subfigure[]{
\centering
\begin{minipage}[t]{0.45\linewidth}
\centering
\includegraphics[width=1.6in]{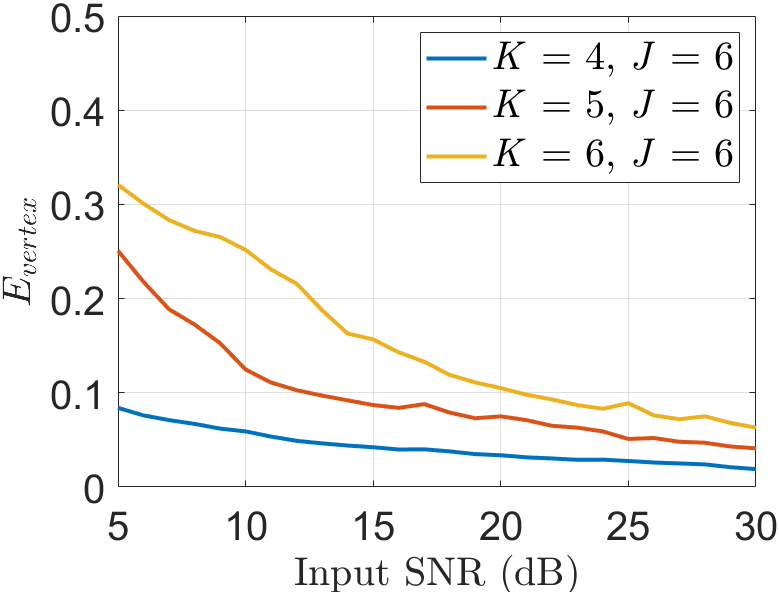}
\end{minipage}%
}%
\subfigure[]{
\centering
\begin{minipage}[t]{0.5\linewidth}
\centering
\includegraphics[width=1.6in]{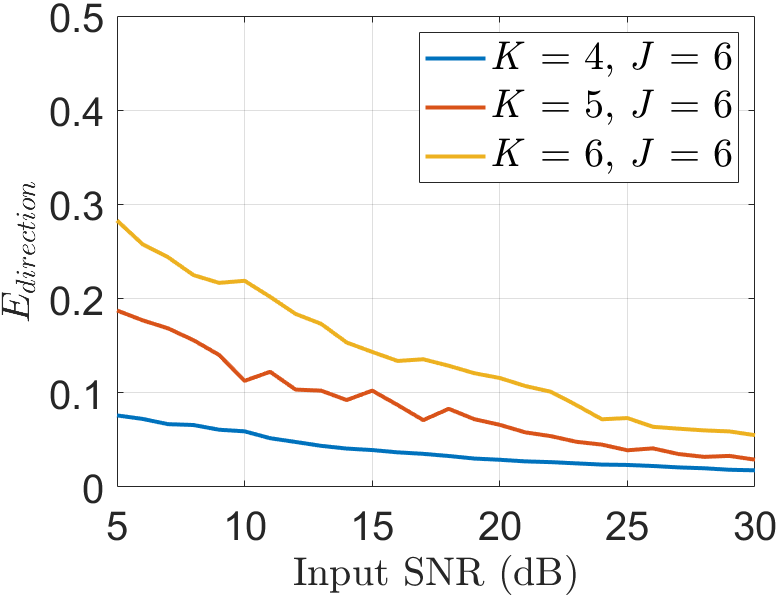}
\end{minipage}%
}%

\centering
\caption{Average reconstruction error. In (a) we show the error of the estimated 3D vertices, and in (b) the error of the estimated direction vectors \textit{wrt.} the SNR level of the 2D samples for $K = 4,\ 5$ and $6$, respectively. The number of projections is $6$.}
\label{fig_different_K}
%\vspace*{-0.5cm}
\end{figure}

\subsection{Noisy Setting}

In this section, we assess the performance of the proposed algorithm by analysing:

\begin{enumerate}
    \item Dependence of the estimation error on the sample noise level.
    \item Dependence of the estimation error on the number of vertices $K$. 
    \item Dependence of the estimation error on the number of projections $J$.
    %\item Towards general sampling kernels.
\end{enumerate}

\begin{figure}[!h]
\centering
\subfigure[]{
  \includegraphics[scale=0.2]{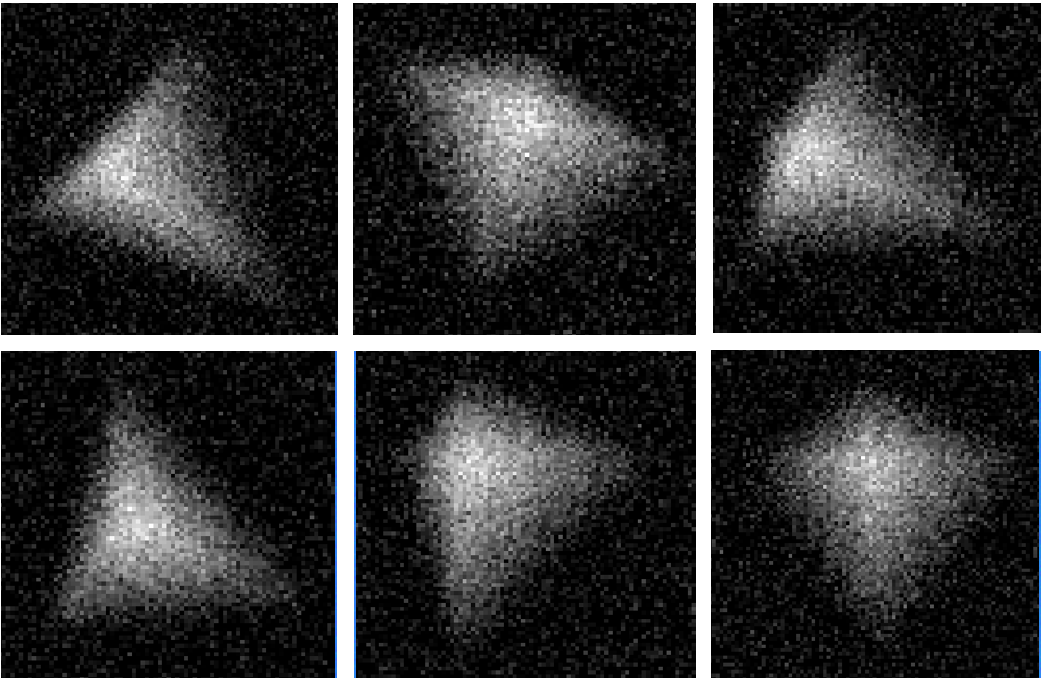}
  }
\subfigure[]{
  \includegraphics[scale=0.2]{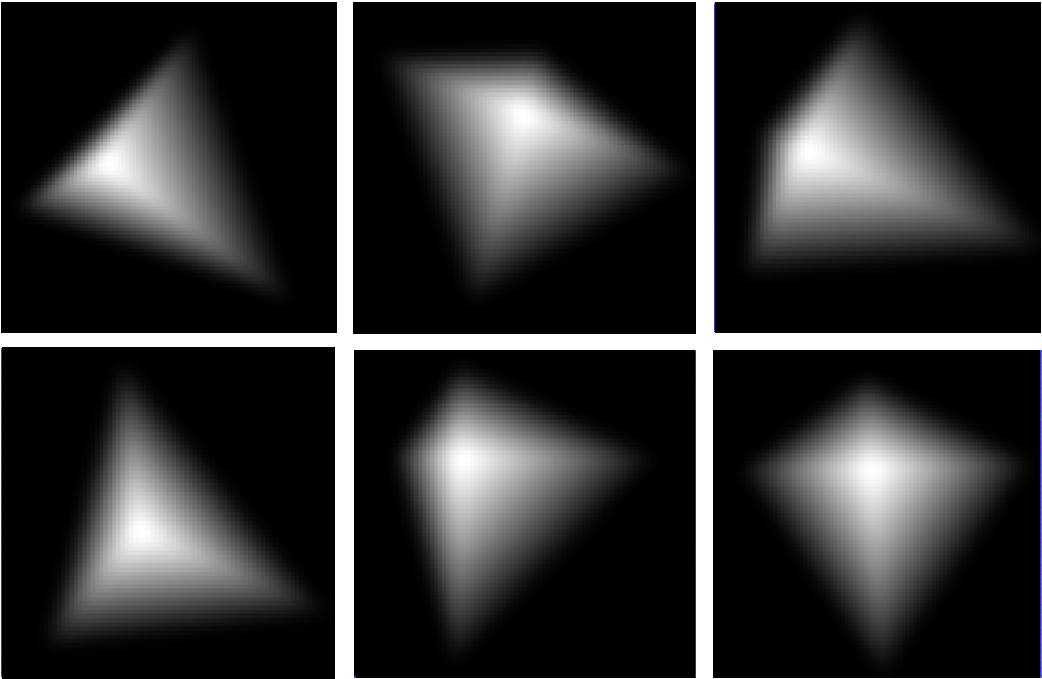}
  }
\subfigure[]{
  \includegraphics[scale=0.2]{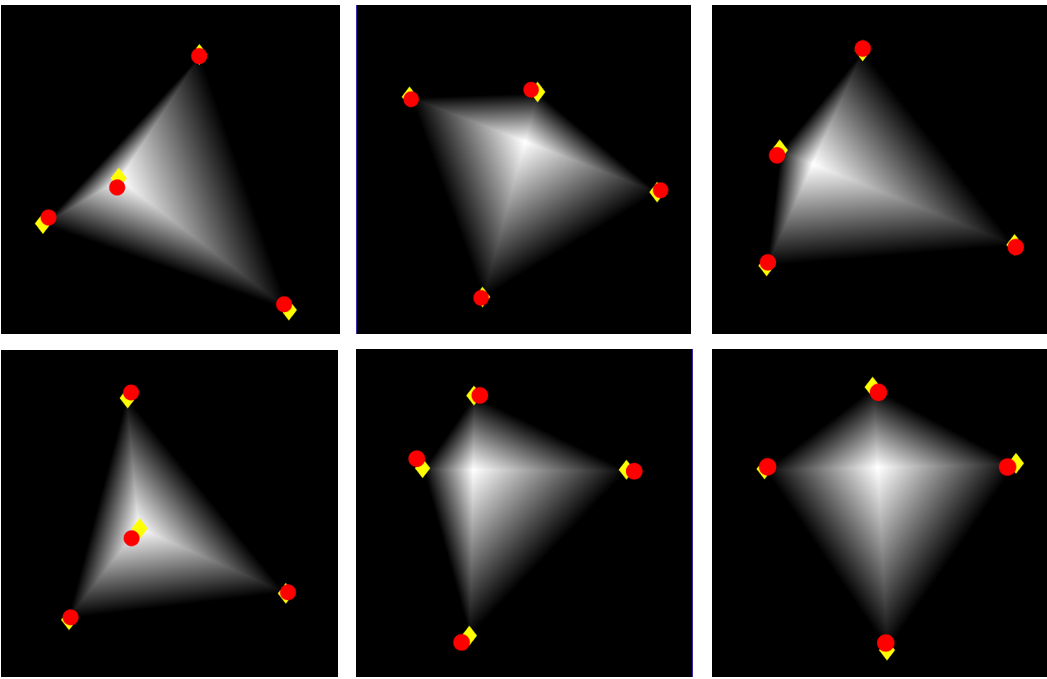}
  }
\subfigure[]{
  \includegraphics[scale=0.2]{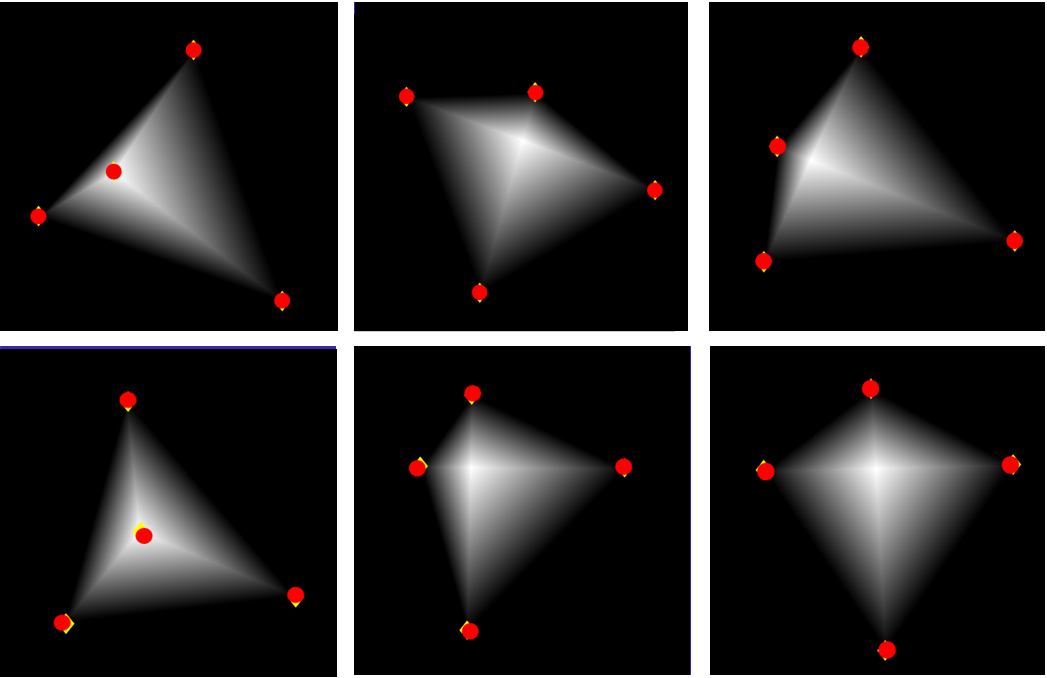}
  }
\subfigure[]{
  \includegraphics[scale=0.23]{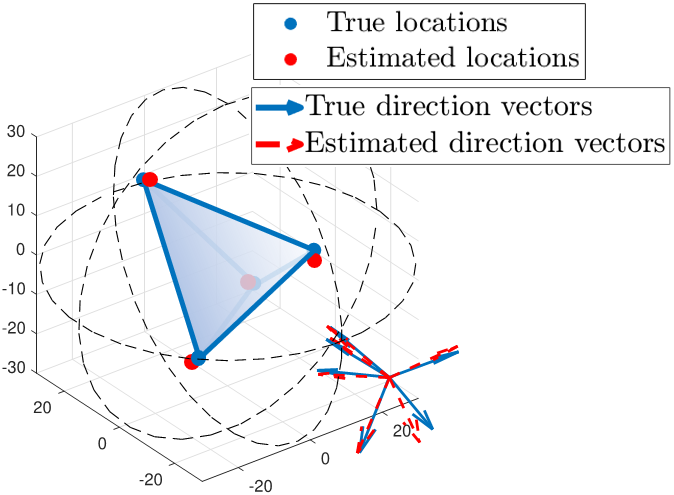}
  }
\subfigure[]{
  \includegraphics[scale=0.23]{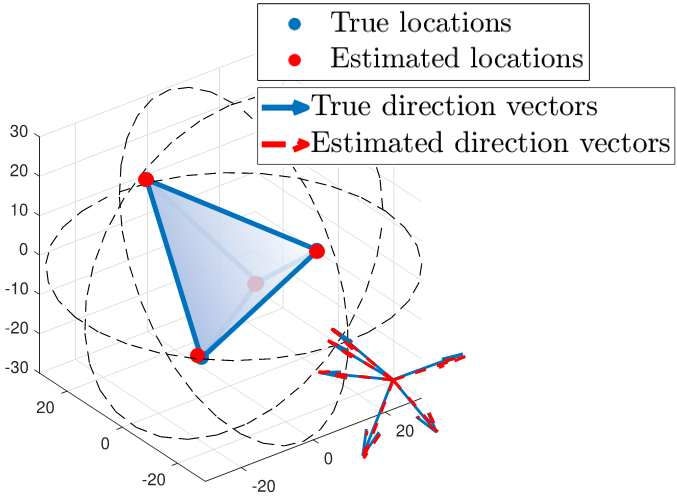}
  }
\centering
\caption{(a) and (b) depict the noisy sampled projections at SNR level of 5 and 30 dB respectively. The 2D locations of the projected vertices are estimated as shown in (c) and (d) respectively. The estimated locations are denoted with red dots while the true locations are denoted with yellow diamonds. (e) and (f) show the final estimation of the 3D polyhedron and direction vectors. The blue dots and arrows represent the true vertices and direction vectors, while the red ones represent the estimated values. For SNR $=5$ dB, the reconstruction errors are $E_{\text{vertex}} = 0.076$ and $E_{\text{direction}} = 0.102$ respectively. When SNR $= 30$ dB, the reconstruction errors are  $E_{\text{vertex}} = 0.018$ and $E_{\text{direction}} = 0.012$ respectively.}
\label{fig_v4}
%\vspace*{-0.4cm}
\end{figure}

\subsubsection{Effect of the noise level}
We consider the case where the samples of the projections are corrupted by additive white Gaussian noise with zero mean and variance $\sigma^{2}$, and the SNR level varies between $5$ dB to $30$ dB. We compare the estimation error for polyhedron models with $K = 4$ to $K = 6$ vertices using $J = 6$ noisy projections. The 2D projections are of size $1024\times1024$ pixels. Their corresponding blurred samples of size $256\times 256$ pixels are obtained by first filtering with a 2D separable B-spline kernel of order $8$ along each axis and downsampling. The coefficients $c_{m,n}^{\omega}$ in Eq.~(\ref{approx_rep}) are generated with $a_{\omega} = 1.2Re^{i2\pi \frac{\omega}{50}}, \omega = 0,1,...,49$. 
For every SNR level, the experiment is repeated $1000$ times on randomly generated polyhedra and direction vectors. 

To achieve robust reconstruction, we apply the algorithm proposed in Section \ref{section4sub1} to estimate the 2D projected locations $\boldsymbol{\hat{p}}_{k,j}$. Moreover, we employ a noise-tolerant pairing strategy based on the rank criterion introduced in Section \ref{section3sub2} and the measurement consistency criterion. The latter criterion requires to reconstruct a test 3D polyhedron model based on the current pairing. It makes use of the fact that given correct pairing, the resynthesized parameters $\boldsymbol{\check{p}}_{k,j}$ from the test model should be consistent with the estimated parameters $\boldsymbol{\hat{p}}_{k,j}$. Namely, the average difference between $
\boldsymbol{\check{p}}_{k,j}$ and $\boldsymbol{\hat{p}}_{k,j}$ should be within a small threshold. We operate by initiating the pairing using the rank criterion, and verify the paired result against the consistency criterion. For a final estimation result, we run the optimization scheme proposed in Section~\ref{section4sub2} using all paired 2D parameters.

\begin{figure}[t]
  \centering
  \subfigure[]{
  \includegraphics[scale=0.28]{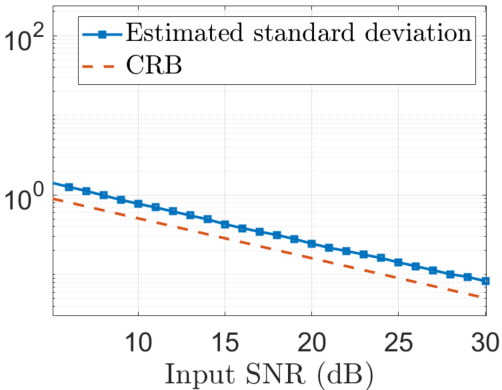}
  }
  %\hspace{0.05in}
  \subfigure[]{
  \includegraphics[scale=0.28]{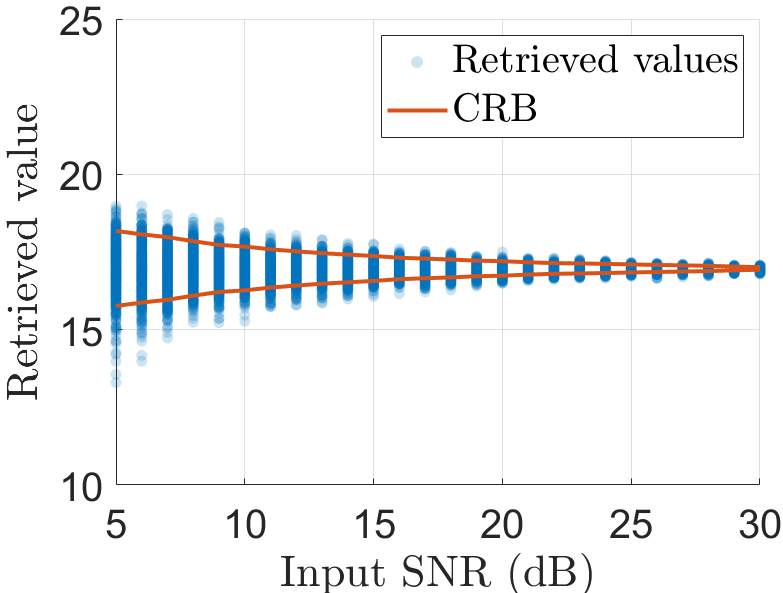}
   }
  \caption{(a) The standard deviation in the estimation of $\boldsymbol{p}_{2,j}^{y}$ and the Cram\'er Rao bound for different levels of noise. (b) Scatter plot of retrieved values of $\boldsymbol{p}_{2,j}^{y}$. }
  \label{crb_bound}
  %\vspace*{-0.4cm}
\end{figure}

Fig.~\ref{fig_different_K} illustrates the average reconstruction error $E_{\text{vertex}}$ and $E_{\text{direction}}$ when the number of vertices $K = 4,5,6$ and the number of projections is $J=6$. We notice that, although 2D samples are quite noisy (SNR = 5 dB), the proposed method can still retrieve accurately the polyhedron and direction vectors.
Fig.~\ref{fig_v4} depicts a visual example where the 2D samples are corrupted with noise at SNR level of $5$ dB and $30$ dB respectively and the number of vertices is $K=4$. 

In addition, we use Cram\'er-Rao bound (CRB) to assess the accuracy of the retrieval of the 2D locations of projected vertices $\boldsymbol{p}_{k,j}$ from the 2D samples. This bound enables us to assess the accuracy of the FRI estimation techniques proposed in Section \ref{section4sub1}. For the setting in Fig.~\ref{crb_bound}, we assume the $K = 4$ projected vertices are located at $\boldsymbol{p}_{1,j} = [-0.75R,0]$, \ $\boldsymbol{p}_{2,j} = [0,0.5R]$,\ $\boldsymbol{p}_{3,j}=[0.75R,0]$ and $\boldsymbol{p}_{4,j}=[0,-0.5R]$.

\subsubsection{Effect of the number of vertices $K$ and number of projections $J$}

The effect of the number $K$ of vertices on the estimation accuracy of a polyhedron is shown in Fig.~\ref{fig_different_K} when we have only access to $J=6$ projections. The figure shows that performance deteriorates with $K$ especially at low SNRs. Theorem~\ref{theorem1} indicates that the minimum number of projection needed for perfect reconstruction is $J = 3$ and is independent of $K$. In practice, one has often access to a number of projections much greater than $J=3$. It is therefore natural to explore the use of this redundancy to improve performance.

First of all, more projections increase the parameter pairing accuracy since there are more parameters to be tested against. Moreover, the linear system in Eq.~(\ref{min_s}) becomes more overdetermined, which yields better estimation results.

\begin{figure}[t]
  \centering
  \subfigure[]{
  \includegraphics[scale=0.29]{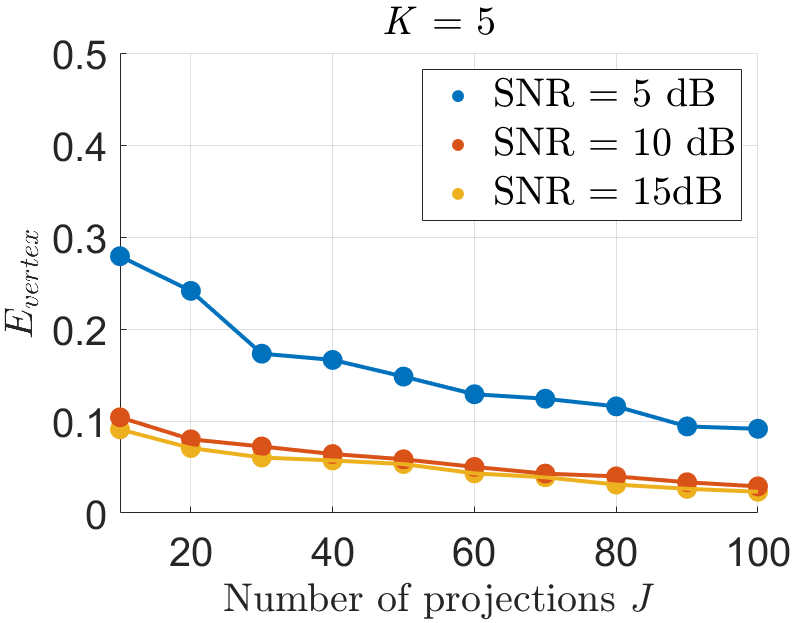}
  }
  %\hspace{0.05in}
  \subfigure[]{
  \includegraphics[scale=0.29]{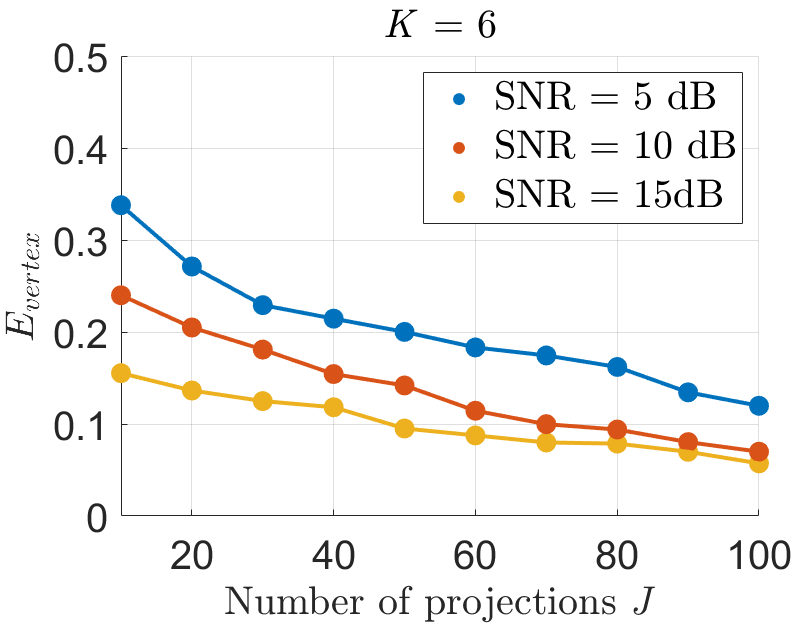}
   }
  \caption{Average reconstruction error $E_{\text{vertex}}$ \emph{wrt.} the number of projections $J$ ranging from $10$ to $100$ for polyhedra with $K=5$ and $6$ vertices respectively.}
  \label{fig_J}
  %\vspace*{-0.4cm}
\end{figure}

Fig.~\ref{fig_J} compares the estimation error \textit{wrt.} the number of projections $J$ when $K=5,6$ and SNR equals to $5$ dB, $10$ dB and $15$ dB respectively. For every $J$, the experiment is repeated $1000$ times. It can be seen that increasing the number of projections yields more accurate estimation results especially when the SNR level is low. For example, it is of interest to note that the performance of our method for $K=5$ when SNR$= 5$ dB and $J=100$ is comparable to the case when $K = 4$ in Fig.~\ref{fig_different_K}. This indicates that we can overcome the performance deterioration due to the complexity of the polyhedron by increasing $J$. A visual example is depicted in Fig.~\ref{rec_v5_new1} where $K = 5$ and the SNR level is $5$ dB for $6$ and $20$ projections respectively. To further justify this claim, we show a visual experiment result in Fig.~\ref{fig_v6} where the polyhedron has $K=6$ vertices and its 2D projection samples are severely corrupted by noise (SNR $= -2$ dB). In order to obtain a faithful reconstruction, we increase the number of projections to $J = 300$. The reconstruction result is shown in Fig.~\ref{fig_v6}~(b). The estimation accuracy is $E_{\text{vertex}}=0.196$ and $E_{\text{direction}}=0.137$, which is comparable to Fig.~\ref{rec_v5_new1}~(c), where $K=5, J= 20$ and SNR$=5$ dB.     

\begin{figure}[t]\label{rec_v5_new1}
  \centering
  \subfigure[]{
  \includegraphics[scale=0.2]{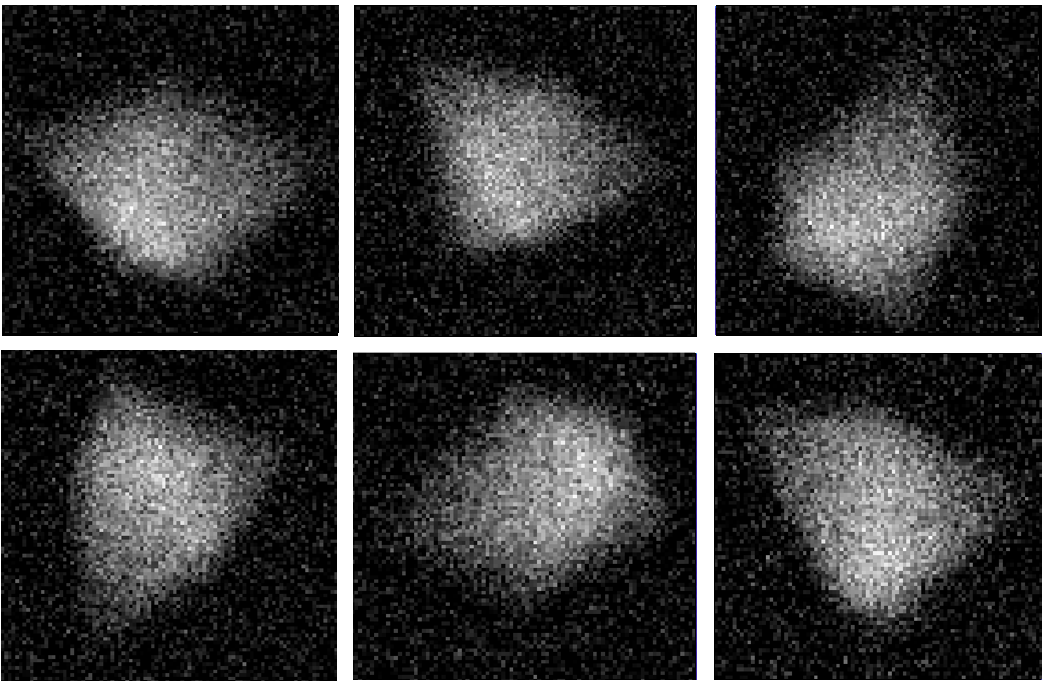}
  }
  %\hspace{0.05in}
  \subfigure[]{
  \includegraphics[scale=0.2]{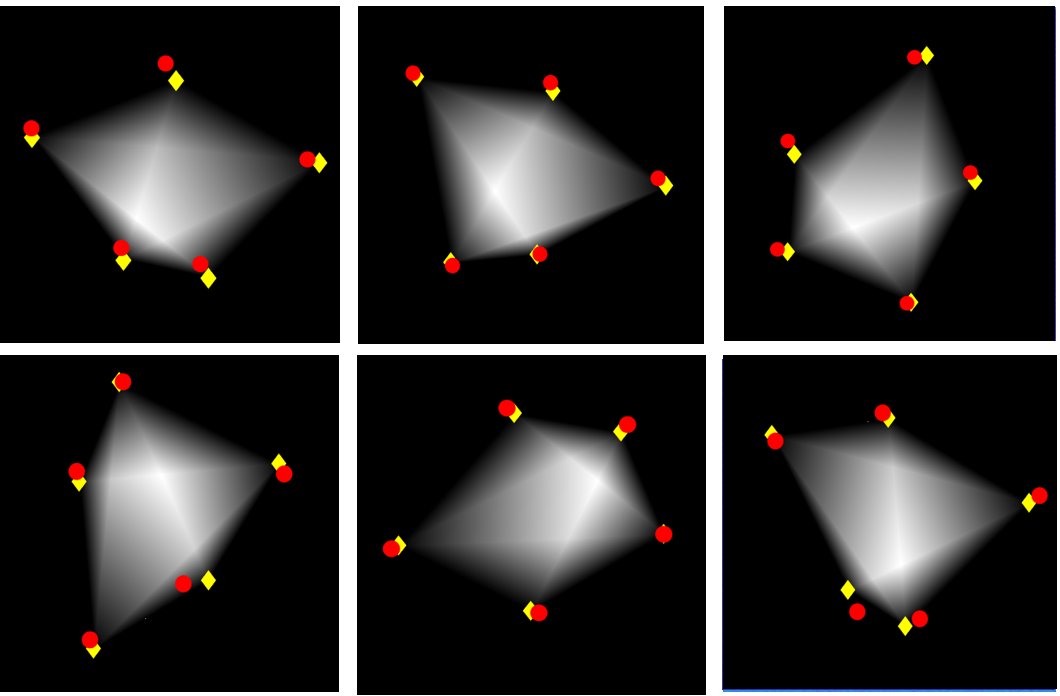}
   }
\subfigure[]{
  \includegraphics[scale=0.245]{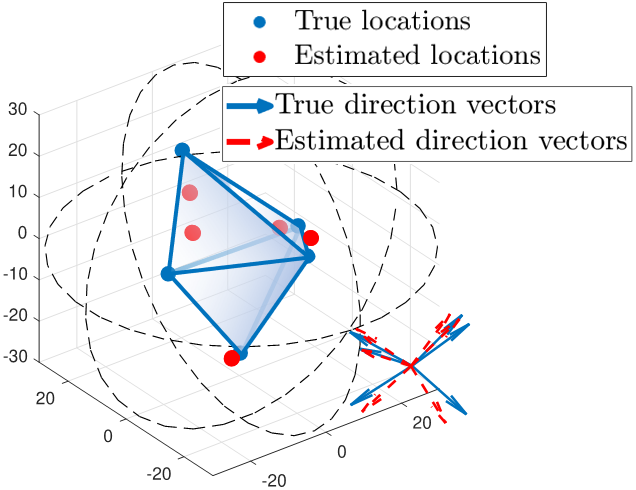}
  }
  \subfigure[]{
  \includegraphics[scale=0.245]{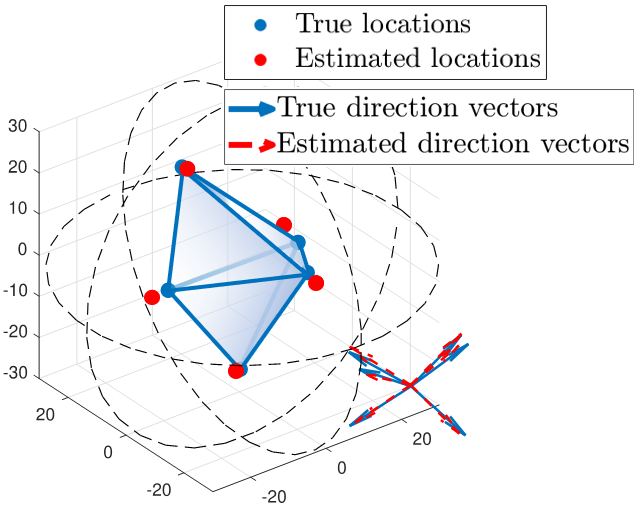}
  }
  \caption{(a) Noisy 2D samples of projections at SNR level of 5dB. (b) Estimation of the 2D locations of the vertices (red dots) compared to the true locations (yellow diamonds). (c) Final estimation result of the polyhedron with $K=5$ vertices from $6$ projections only. $E_{\text{vertex}} = 0.392$ and $E_{\text{direction}} = 0.235$. (d) Final estimation result of the same polyhedron using $20$ projections. $E_{\text{vertex}} = 0.187$ and $E_{\text{direction}} = 0.116$.}
  \label{rec_v5_new1}
  %\vspace*{-0.4cm}
\end{figure}

\begin{figure}[!t]
\subfigure[]{
\centering
\begin{minipage}[t]{0.45\linewidth}
\centering
\includegraphics[width=1.45in]{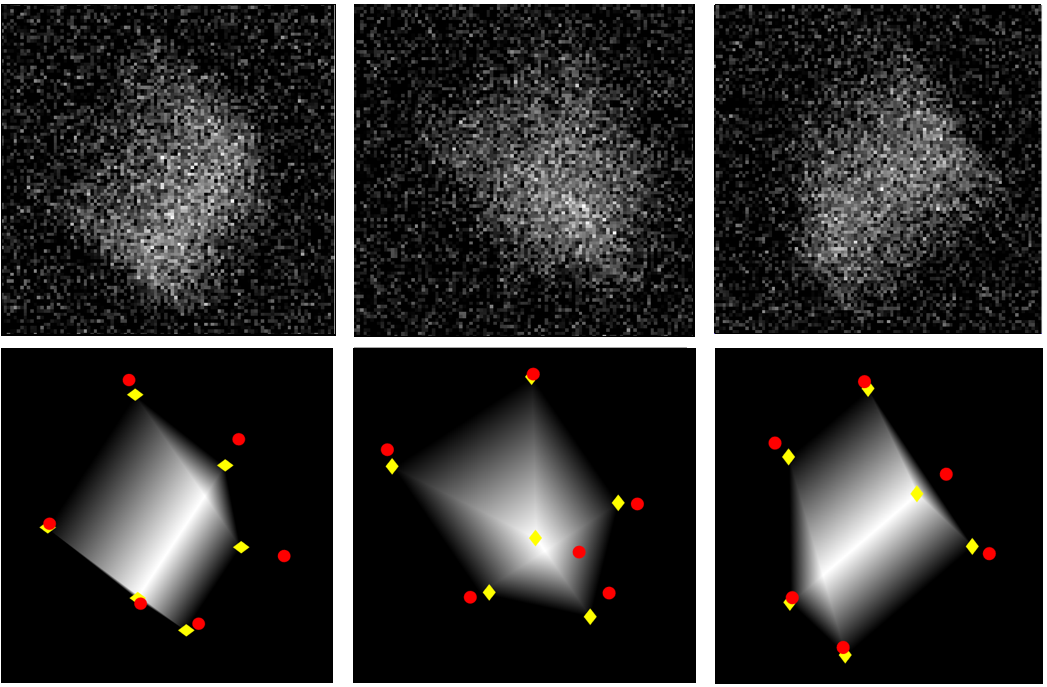}
\end{minipage}%
}%
\subfigure[]{
\centering
\begin{minipage}[t]{0.5\linewidth}
\centering
\includegraphics[width=1.4in]{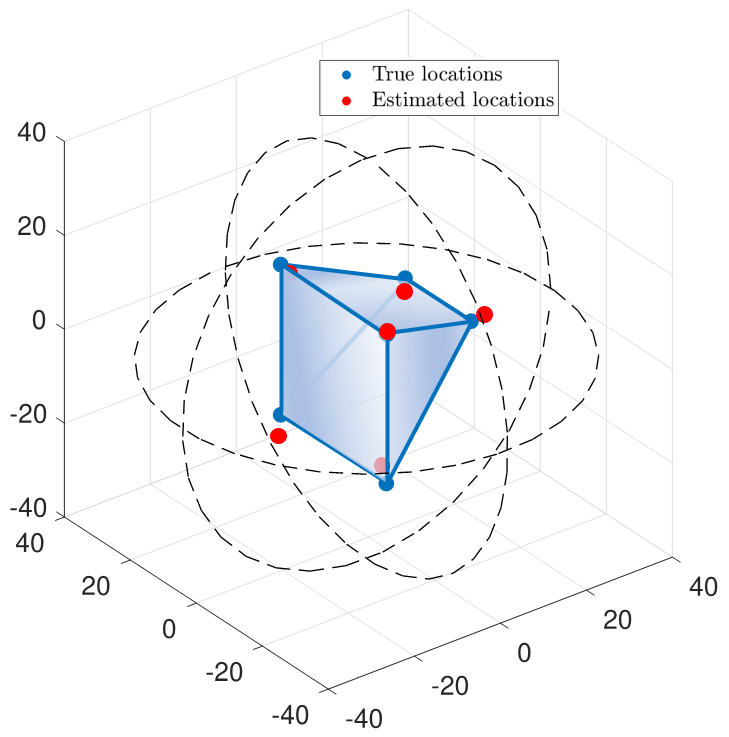}
\end{minipage}%
}%

\centering
\caption{(a) Top row figures show examples of the noisy projection samples corrupted with additive Gaussian noise at SNR = $-2$ dB. While bottom row figures show the estimation of the 2D locations (red dots) of the vertices compared to the true vertices (yellow diamonds). (b) Estimation result of the polyhedron with $K = 6$ projections and $J=300$ projections. $E_{\text{vertex}}=0.196$ and $E_{\text{direction}}=0.137$.}
\label{fig_v6}
%\vspace*{-0.4cm}
\end{figure}

\subsection{Real Computed Tomography Data}\label{section6}

We apply our proposed method to the geometrical calibration of real CT scanning data, obtained at the FleX-ray laboratory at the Center for Mathematics \& Computer Science, Amsterdam. Geometrical calibration is a critical step towards ensuring traceability in CT measurement. Often during the calibration process, a geometrically patterned reference object is imaged, then the system geometry, i.e. the projection angles, is deduced from the 2D projection measurement (radiographs). For this purpose, our proposed framework provides a theoretical guide and a practical tool to perform the calibration. Specifically, as a first step, one should extract paired projected locations of the reference objects. Then, the proposed robust algorithm in Section~\ref{section4sub2} can be applied for the estimation of the 3D system geometry. In what follows, we explain the experiment in details.

\begin{figure}[t]
  \centering
  \subfigure[]{
  \includegraphics[scale=0.3]{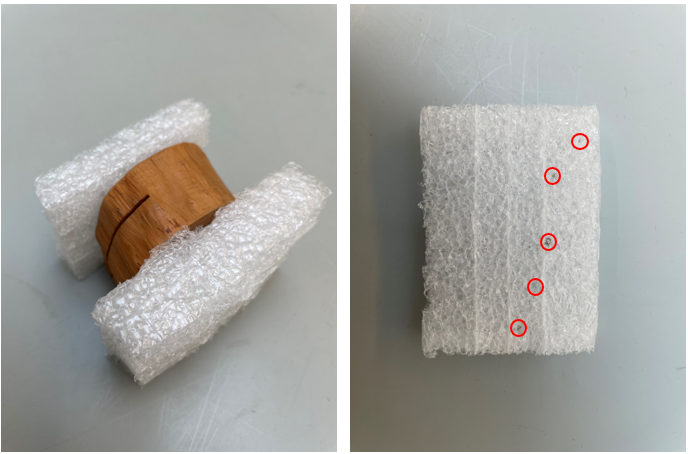}
  }
  \subfigure[]{
  \includegraphics[scale=0.3]{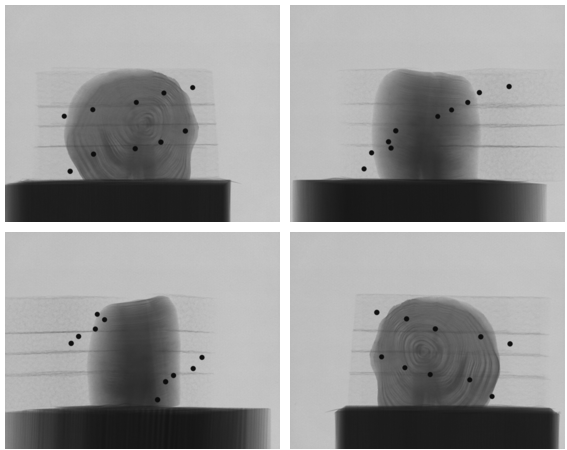}
  }
  %\hspace{0.05in}
  \subfigure[]{
  \includegraphics[scale=0.33]{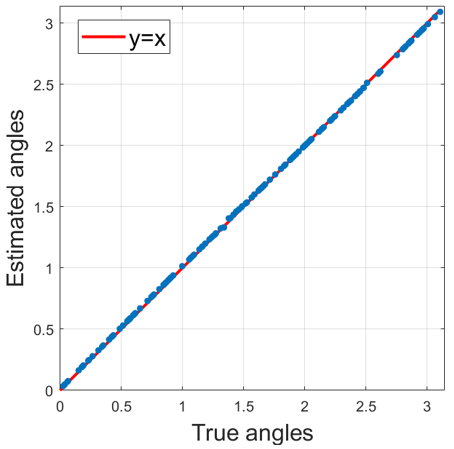}
   }
   \ \ \ \ \ \
\subfigure[]{
  \includegraphics[scale=0.33]{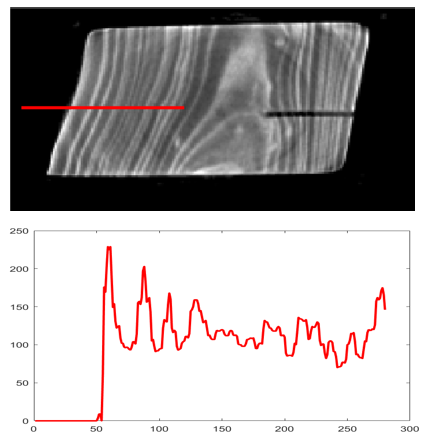}
  }
  \caption{(a) The wooden block is surrounded by the foam, inside which the markers (red circle) are placed. (b) 2D radiographs of the wooden block with markers. (c) Estimation of the 2D projection angles vs. 2D angles given by the FleX-ray laboratory. (d) Reconstruction of a slice of the wooden block using the estimated angles.}
  \label{realdataproj}
  %\vspace*{-0.6cm}
\end{figure}

In the experiment, the projection measurements contain $J = 200$ 2D radiographs of a wooden block, surrounded by two pieces of foam in which $10$ small metal balls (markers) are placed, as shown in Fig.~\ref{realdataproj}~(a). The metal balls serve as reference objects, facilitating the geometrical calibration of the scanning system, specifically, to estimate the projection angles. The radiographs are obtained using a micro-CT scanner. During the scanning process, the object is placed on a stable rotation stage, and subsequently rotates the object between a static source and a detector. In other words, the angular motion is only on a plane. The detector plane is placed vertically to the ground. Therefore, the projection directions $\{\boldsymbol{d}_{j}\}_{j=1}^{J}$ are co-planar on the horizontal plane. Moreover, the unit direction vectors $\{\boldsymbol{u}_{j,x}\}_{j=1}^{J}$ are the same across all projections, and they assume the value $[0,0,1]^{T}$. In Fig.~\ref{realdataproj}~(b), examples of the 2D radiographs at unknown projection angles are shown.

From the 2D radiographs taken at unknown angles, we aim to estimate the projection angles. We first find the projected marker locations $\boldsymbol{p}_{k,j}$ on each 2D radiograph, from which we extract the projection directions $\boldsymbol{d}_{j}$ using the method presented in Section \ref{section3sub3}. Since the direction vectors $\{\boldsymbol{u}_{j,x}\}_{j=1}^{J}$ are the same across all projections, given the 2D projected locations
of the markers, the estimation of the projected directions is
actually a degenerate case of our 3D estimation method, and
we can directly apply a 2D method similar to the one that we
proposed in \cite{me} using $p_{k,j}^{y}$ only. We plot the estimated angles using our proposed method against the 2D angles given by the metadata of the scanner in Fig.~\ref{realdataproj} (c). The average angle estimation error is $0.005$~rad and the standard deviation for the estimation of angles is $0.013$~rad.

For illustrative reasons, we also reconstruct a slice of the wooden block using the FISTA SWLS-TV algorithm provided by the ToMoBAR toolbox~\cite{tomobar} with the estimated angles. The reconstruction result is shown in Fig.~\ref{realdataproj} (d). 

\section{Conclusion} \label{section7}
In this paper, we have addressed the sampling problem of estimating an unknown 3D polyhedron from a small number of sampled 2D tomographic projections at unknown angles. The method is able to estimate exactly the projection directions and the 3D structure up to an orthogonal transformation. Interestingly, we are able to show that the minimum number of projections required is unrelated to the complexity of the 3D object. From various experiments, the algorithm is shown to achieve accurate estimation even when the samples are corrupted with noise. Moreover, experimental results on real data show the potential of the proposed approach.

\appendices

\section{Proof of LEMMA1}
\label{proof_lemma1}
\begin{proof}
Let $L=\iint_{\Pi_{j}} I_{j}(x,y)h'''(z)\ dxdy$. By substituting $I_{j}(x,y)$ with the expression in Eq.~(\ref{eq3}), $L$ can be written as:
\begin{align}
\label{eq1_append_new}
   \nonumber L &=\iint_{\Pi_{j}} \int_{\mathbb{R}^{3}}g(\boldsymbol{r})\delta\left(x-\boldsymbol{u}_{j,x}^{T}\boldsymbol{r}-s_{j,x}\right)\\
   \nonumber &\cdot\delta\left(y-\boldsymbol{u}_{j,y}^{T}\boldsymbol{r}-s_{j,y}\right)d^{3}\boldsymbol{r}h'''(z)\ dxdy\\
   \nonumber &\overset{(a)}{=}\int_{\mathbb{R}^{3}}g(\boldsymbol{r})h'''\left(\boldsymbol{u}_{j,x}^{T}\boldsymbol{r}+s_{j,x}+i\left(\boldsymbol{u}_{j,y}^{T}\boldsymbol{r}+s_{j,y}\right)\right)d^{3}\boldsymbol{r}\\
   \nonumber &=\int_{\mathbb{R}^{3}}g(\boldsymbol{r})h'''\left({\underbrace{\left(\boldsymbol{u}_{j,x}+i\boldsymbol{u}_{j,y}\right)}_{\boldsymbol{u}_{j}}}^{T}\boldsymbol{r}+\underbrace{\left(s_{j,x}+is_{j,y}\right)}_{s_{j}}\right)d^{3}\boldsymbol{r}  \\
   &= \int_{\mathbb{R}^{3}}g(\boldsymbol{r})h'''(\boldsymbol{u}_{j}^{T}\boldsymbol{r}+s_{j})d^{3}\boldsymbol{r},
\end{align}
where $\boldsymbol{u}_{j}\in\mathbb{C}^{3}$, $s_{j}\in\mathbb{C}$, and $(a)$ follows from the property of the Dirac Delta function $\delta(\cdot)$.

For any vector $\boldsymbol{w}\in\mathbb{C}^{3}$, the following equality holds:
\begin{equation}\label{eq2_append_new}
    \Div\left(h''(\boldsymbol{u}_{j}^{T}\boldsymbol{r}+s_{j})\boldsymbol{w}\right)=h'''\left(\boldsymbol{u}_{j}^{T}\boldsymbol{r}+s_{j}\right)\boldsymbol{u}_{j}^{T}\boldsymbol{w},
\end{equation}
where $\Div$ denotes divergence.
Therefore, if we multiply both sides in Eq.~(\ref{eq1_append_new}) with $\boldsymbol{u}_{j}^{T}\boldsymbol{w}$, it yields:
\begin{align}
\label{eq3_append_new}
    \nonumber \boldsymbol{u}_{j}^{T}\boldsymbol{w}L&=\int_{\mathbb{R}^{3}}g(\boldsymbol{r})h'''\left(\boldsymbol{u}_{j}^{T}\boldsymbol{r}+s_{j}\right)\boldsymbol{u}_{j}^{T}\boldsymbol{w}d^{3}\boldsymbol{r}\\
    \nonumber&\overset{(b)}{=}\int_{\mathbb{R}^{3}}g(\boldsymbol{r})\Div\left(h''\left(\boldsymbol{u}_{j}^{T}\boldsymbol{r}+s_{j}\right)\boldsymbol{w}\right)d^{3}\boldsymbol{r}\\
    \nonumber &\overset{(c)}{=}\int_{\Gamma}\Div\left(h''\left(\boldsymbol{u}_{j}^{T}\boldsymbol{r}+s_{j}\right)\boldsymbol{w}\right)d^{3}\boldsymbol{r}\\
    &\overset{(d)}{=}\oiint_{\partial\Gamma} h''\left(\boldsymbol{u}_{j}^{T}\boldsymbol{r}+s_{j}\right)\boldsymbol{w}^{T}d\boldsymbol{S}.
\end{align}
where $(b)$ follows from Eq.~(\ref{eq2_append_new}), $(c)$ follows from Eq.~(\ref{eq1}) and $(d)$ follows from the divergence theorem.

We note that the surface of a polyhedron $\partial \Gamma$ is made of flat polygonal faces, and every such polygonal face can be decomposed into disjoint triangular areas whose vertices are the vertices of the polyhedron. Please see Fig.~\ref{fig1_append1} for an example.
\begin{comment}
For clarity, let $k,l,m\in [1,K]$ denote distinct integers, and we define the set of all possible trios that identify a triangular face as follows:
\begin{equation*}
    \mathcal{S}_{\Gamma} =\{(k,l,m):
\triangle{\boldsymbol{v}_{k}\boldsymbol{v}_{l}\boldsymbol{v}_{m}} \subset \partial \Gamma\}, 
\end{equation*}
where $\triangle{\boldsymbol{v}_{k}\boldsymbol{v}_{l}\boldsymbol{v}_{m}}$ denotes the triangular face whose vertices are $\boldsymbol{v}_{k}$, $\boldsymbol{v}_{l}$ and $\boldsymbol{v}_{m}$. For example, in Fig.~\ref{fig1_append1}, $(k,l,m)$, $(k,l,n)$, $(k,m,n)$ and $(l,m,n)$ all belong to the set $\mathcal{S}_{\Gamma}$. Furthermore, we denote the irreducible set of trios of a polyhedron surface $\partial \Gamma$ as $\mathcal{T}_{\Gamma}$, $\mathcal{T}_{\Gamma}\subseteq \mathcal{S}_{\Gamma}$,  and it satisfies:
\begin{equation*} \bigcup_{t_{n}\in\mathcal{T}_{\Gamma}}t_{n}=\partial \Gamma  \text{ and } \forall t_{p}\in\mathcal{T}_{\Gamma}, \left(\bigcup_{t_{n}\in\mathcal{T}_{\Gamma}}t_{n}\right) \setminus t_{p}\subset \partial\Gamma.
\end{equation*}
A polyhedron surface $\partial \Gamma$ can induce more than one irreducible set, as long as the union of the triangular area indexed by all trios in the irreducible set forms exactly the surface area $\partial \Gamma$. Take Fig.~\ref{fig1_append1} as an example, even though $(k,l,m)$ and $(l,m,n)$ both belongs to $\mathcal{S}_{\Gamma}$, they cannot both belong to $\mathcal{T}_{\Gamma}$.
\end{comment}

Let $k,l,m\in\mathcal{T}_{\Gamma}$ be distinct integers. We denote with $\triangle\boldsymbol{v}_{k}\boldsymbol{v}_{l}\boldsymbol{v}_{m}$ the triangular surface whose vertices are $\boldsymbol{v}_{k}$, $\boldsymbol{v}_{l}$ and $\boldsymbol{v}_{m}$, and $\mathcal{T}_{\Gamma}$ represents the set of triplets $k,l.m$ such that all triangular faces forming $\partial \Gamma$ are included. The closed surface integral in Eq. (\ref{eq3_append_new}) can therefore be written as a sum of integral over triangular faces as follows:
\begin{align}
\label{eq4_append_new}
\nonumber\boldsymbol{u}_{j}^{T}\boldsymbol{w}L&=\oiint_{\partial \Gamma}h''\left(\boldsymbol{u}_{j}^{T}\boldsymbol{r}+s_{j}\right)\boldsymbol{w}^{T}d\boldsymbol{S}\\
&=\sum_{k,l,m\in\mathcal{T}_{\Gamma}}\ \ \ \ 
\iint\limits_{\mathclap{\triangle{\boldsymbol{v}_{k}\boldsymbol{v}_{l}\boldsymbol{v}_{m}}}}
h''\left(\boldsymbol{u}_{j}^{T}\boldsymbol{r}+s_{j}\right)\boldsymbol{w}^{T}d\boldsymbol{S}_{k,l,m},
\end{align}
where $d\boldsymbol{S}_{k,l,m}=\boldsymbol{n}_{k,l,m}dS$, and $\boldsymbol{n}_{k,l,m}\in\mathbb{R}^{3}$ is the outward pointing unit vector normal to the triangular face $\triangle{\boldsymbol{v}_{k}\boldsymbol{v}_{l}\boldsymbol{v}_{m}}$, as shown in Fig.~\ref{fig1_append1}.

We first consider the integral on a single triangular face $\triangle{\boldsymbol{v}_{k}\boldsymbol{v}_{l}\boldsymbol{v}_{m}}$. For $\forall \boldsymbol{r}\in\triangle{\boldsymbol{v}_{k}\boldsymbol{v}_{l}\boldsymbol{v}_{m}}$, $\boldsymbol{r}$ can be represented as:
\begin{equation*}
    \boldsymbol{r}=\boldsymbol{v}_{k}+\lambda(\boldsymbol{v}_{l}-\boldsymbol{v}_{k}) +\mu(\boldsymbol{v}_{m}-\boldsymbol{v}_{k}),
\end{equation*}
where $\lambda,\mu\in\mathbb{R}$ and $0\leq \lambda,\mu,\lambda+\mu\leq 1$. Therefore, 
\begin{align}
\label{eq5_append_new}
    \nonumber d\boldsymbol{S}_{k,l,m}&=\frac{\partial \boldsymbol{r}}{\partial \lambda}\times \frac{\partial \boldsymbol{r}}{\partial \mu}d\lambda d\mu\\
    &=\underbrace{\left(\boldsymbol{v}_{l}-\boldsymbol{v}_{k}\right)\times \left(\boldsymbol{v}_{m}-\boldsymbol{v}_{k}\right)}_{\boldsymbol{v}_{k,l,m}}d\lambda d \mu.
\end{align}
The integral over a single triangular face can then be written as:
\begin{align}
\label{eq6_append_new}
\nonumber&\iint\limits_{\mathclap{\triangle{\boldsymbol{v}_{k}\boldsymbol{v}_{l}\boldsymbol{v}_{m}}}}h''\left(\boldsymbol{u}_{j}^{T}\boldsymbol{r}+s_{j}\right)\boldsymbol{w}^{T}d\boldsymbol{S}_{k,l,m}\\
\nonumber=&\iint\limits_{\substack{0\leq \lambda,\mu\leq 1\\ \lambda+\mu\leq 1}}h''\left(\boldsymbol{u}_{j}^{T}(\boldsymbol{v}_{k}+\lambda(\boldsymbol{v}_{l}-\boldsymbol{v}_{k})+\mu(\boldsymbol{v}_{m}-\boldsymbol{v}_{k}))+s_{j}\right)\\
\nonumber&\cdot\boldsymbol{w}^{T}\boldsymbol{v}_{k,l,m}\ d\lambda d\mu\\
\nonumber= &\ \boldsymbol{w}^{T}\boldsymbol{v}_{k,l,m}\Biggl(\frac{h(\boldsymbol{u}_{j}^{T}\boldsymbol{v}_{k}+s_{j})}{\boldsymbol{u}_{j}^{T}\left(\boldsymbol{v}_{k}-\boldsymbol{v}_{l}\right)\boldsymbol{u}_{j}^{T}\left(\boldsymbol{v}_{k}-\boldsymbol{v}_{m}\right)}\\
\nonumber &\quad\quad\quad\quad+\frac{h(\boldsymbol{u}_{j}^{T}\boldsymbol{v}_{l}+s_{j})}{\boldsymbol{u}_{j}^{T}\left(\boldsymbol{v}_{l}-\boldsymbol{v}_{k}\right)\boldsymbol{u}_{j}^{T}\left(\boldsymbol{v}_{l}-\boldsymbol{v}_{m}\right)}\\
&\quad\quad\quad\quad+\frac{h(\boldsymbol{u}_{j}^{T}\boldsymbol{v}_{m}+s_{j})}{\boldsymbol{u}_{j}^{T}\left(\boldsymbol{v}_{m}-\boldsymbol{v}_{k}\right)\boldsymbol{u}_{j}^{T}\left(\boldsymbol{v}_{m}-\boldsymbol{v}_{l}\right)}\Biggr).
\end{align}
For the above equation to be meaningful, we require that $\forall k\neq l,\ \boldsymbol{u}_{j}^{T}(\boldsymbol{v}_{k}-\boldsymbol{v}_{l})\neq 0$. This is coherent with our assumption that any two projected vertices should not coincide with each other. Then, by letting $\boldsymbol{w}=\boldsymbol{u}^*_{j}$ and substituting Eq.~(\ref{eq6_append_new}) into Eq.~(\ref{eq4_append_new}), we obtain that:
\begin{align}
\label{eq7_append_new}
\nonumber L&=\sum\limits_{\mathclap{{k,l,m}}} \frac{\boldsymbol{u}^{\dag}_{j}\boldsymbol{v}_{k,l,m}}{2}\Biggl(\frac{h(\boldsymbol{u}_{j}^{T}\boldsymbol{v}_{k}+s_{j})}{\boldsymbol{u}_{j}^{T}\left(\boldsymbol{v}_{k}-\boldsymbol{v}_{l}\right)\boldsymbol{u}_{j}^{T}\left(\boldsymbol{v}_{k}-\boldsymbol{v}_{m}\right)}\\
\nonumber \nonumber&+\frac{h(\boldsymbol{u}_{j}^{T}\boldsymbol{v}_{l}+s_{j})}{\boldsymbol{u}_{j}^{T}\left(\boldsymbol{v}_{l}-\boldsymbol{v}_{k}\right)\boldsymbol{u}_{j}^{T}\left(\boldsymbol{v}_{l}-\boldsymbol{v}_{m}\right)}\\
\nonumber&+\frac{h(\boldsymbol{u}_{j}^{T}\boldsymbol{v}_{m}+s_{j})}{\boldsymbol{u}_{j}^{T}\left(\boldsymbol{v}_{m}-\boldsymbol{v}_{k}\right)\boldsymbol{u}_{j}^{T}\left(\boldsymbol{v}_{m}-\boldsymbol{v}_{l}\right)}\Biggr)\\
\nonumber&\overset{(e)}{=}\sum\limits_{\mathclap{{k,l,m}}} \frac{\boldsymbol{u}^{\dag}_{j}\boldsymbol{v}_{k,l,m}}{2}\Biggl(\frac{h(z_{k,j})}{\boldsymbol{u}_{j}^{T}\left(\boldsymbol{v}_{k}-\boldsymbol{v}_{l}\right)\boldsymbol{u}_{j}^{T}\left(\boldsymbol{v}_{k}-\boldsymbol{v}_{m}\right)}\\
\nonumber\nonumber &+\frac{h(z_{l,j})}{\boldsymbol{u}_{j}^{T}\left(\boldsymbol{v}_{l}-\boldsymbol{v}_{k}\right)\boldsymbol{u}_{j}^{T}\left(\boldsymbol{v}_{l}-\boldsymbol{v}_{m}\right)}\\
&+\frac{h(z_{m,j})}{\boldsymbol{u}_{j}^{T}\left(\boldsymbol{v}_{m}-\boldsymbol{v}_{k}\right)\boldsymbol{u}_{j}^{T}\left(\boldsymbol{v}_{m}-\boldsymbol{v}_{l}\right)}\Biggr),
\end{align}
where $(e)$ follows from 
\begin{equation*}
\boldsymbol{u}_{j}^{T}\boldsymbol{v}_{k}+s_{j}=\boldsymbol{u}_{j,x}^{T}\boldsymbol{v}_{k}+s_{j,x}+i\left(\boldsymbol{u}_{j,y}^{T}\boldsymbol{v}_{k}+s_{j,y}\right) =z_{k,j}.
\end{equation*}
If we consider the contribution from each projected vertex $z_{k,j}$ in Eq.~(\ref{eq7_append_new}), then $L$ can be expressed alternatively as:
\begin{align*}
L&=\sum_{k,l,m}\frac{\boldsymbol{u}^{\dag}_{j}\boldsymbol{v}_{k,l,m}h(z_{k,j})}{2\boldsymbol{u}_{j}^{T}\left(\boldsymbol{v}_{k}-\boldsymbol{v}_{l}\right)\boldsymbol{u}_{j}^{T}\left(\boldsymbol{v}_{k}-\boldsymbol{v}_{m}\right)}\\
&=\sum_{k}\rho_{k}h(z_{k,j}),
\end{align*}
which concludes the proof.

\begin{figure}[t]
\centering
\includegraphics[width=0.2\textwidth]{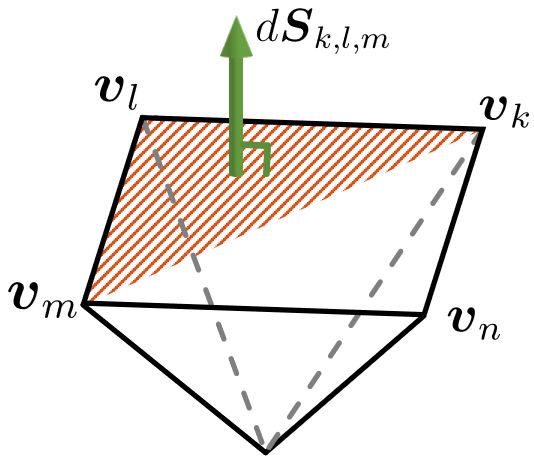}
\caption{The surface of a polyhedron is made of disjoint triangular faces. }
\label{fig1_append1}
%\vspace*{-0.4cm}
\end{figure}

\end{proof}

\section{Annihilating filter based method for retrieving the 2D parameters}
\label{annihilating_filter}
From Eq.~(\ref{approx_sig_mom}) and Eq.~(\ref{eq18}), the exact values of $\eta_{\omega}$ can be written as the ratio of two polynomials:
\begin{align*}
    \eta_{\omega}&=\sum_{k}\frac{\rho_{k}}{z_{k,j}-a_{\omega}}\\
    &\overset{(a)}{=} \frac{\sum_{k}a_{0}\lambda_{k}e^{i\frac{2\pi}{W}k\omega}}{Q_{K}(a_{\omega})}=\frac{\sum_{k}\tilde{\lambda}_{k}e^{i\alpha k \omega}}{Q_{K}(a_{\omega})}=\frac{r_{\omega}}{Q_{K}(a_{\omega})} ,
\end{align*}
where (a) comes from $a_{\omega} = a_{0}e^{i2\pi\frac{\omega}{W}}$ and $\tilde{\lambda}_{k} = a_{0}\lambda_{k}$. The goal is to retrieve $z_{k,j}$ from $\{\eta_{\omega}\}_{\omega=0}^{W}$. To achieve this, we consider a discrete filter $f = \{f_{k}\}_{k = 0}^{K}$ whose zeros are located at $e^{ik\alpha }$ for $k = 0,...,K-1$, in other words:
\begin{equation*}
    F(z) = \sum_{k}f_{k}z^{-1}=\prod_{k}\left(1-e^{ik\alpha}z^{-1} \right).
\end{equation*}
Consequently, the discrete filter also annihilates the sequence $r_{\omega}$, namely $f*r=0$, which can be proven as follows:
\begin{align*}
    \{f*r\}_{\omega}&=\sum_{l}f_{l}r_{\omega-l}\\
    &=\sum_{l}f_{l}\sum_{m}\tilde{\lambda}_{m}e^{i\alpha m (\omega-l)}\\
    &=\sum_{m}\tilde{\lambda}_{m}\sum_{l}f_{l}e^{i\alpha m (\omega-l)}\\
    &=\sum_{m}\tilde{\lambda}_{m}e^{i\alpha m \omega}\sum_{l}f_{l}e^{-i\alpha ml}\\
    &=\sum_{m}\tilde{\lambda}_{m}e^{i\alpha m\omega}F(e^{i\alpha m})=0.
\end{align*}
The above annihilation relation can be further written as:
\begin{align}\label{eq46}
    \nonumber\{f*r\}_{\omega} &= \sum_{l}f_{\omega-l}r_{l}=\sum_{l}f_{\omega-l}Q_{K}(a_{l})\eta_{l}\\
    \nonumber&=\sum_{l}f_{\omega-l}\sum_{k}q_{k}a_{l}^{k}\eta_{l}\\
    &=\sum_{k}q_{k}\underbrace{\sum_{l}f_{\omega-l}a_{l}^{k}\eta_{l}}_{A_{\omega,k}}=0,
\end{align}
where $q_{k}$ denotes the coefficients of the polynomial $Q_{K}$. Therefore, Eq.~(\ref{eq46}) represents another annihilation relation:
\begin{equation}\label{eq47}
    \sum_{k}A_{\omega,k}q_{k} = 0,
\end{equation}
which can be written in the matrix form $\mathbf{A}\mathbf{q} = \boldsymbol{0}$, where $\mathbf{q} = [q_{0},\hdots,q_{K}]^{T}$ and $\mathbf{A} = \mathbf{F}\boldsymbol{\eta}\mathbf{a}$, with:
\begin{equation*}
    \mathbf{F} = 
    \begin{bmatrix}
        f_{K}&\hdots&f_{0}&0&\hdots&0\\
        0&f_{K}&\hdots&f_{0}&\hdots&0\\
        \vdots&\ddots&\ddots&\vdots&\ddots&\vdots\\
        0&\hdots&0&f_{K}&\hdots&f_{0}
    \end{bmatrix}_{(W-K)\times W},
\end{equation*}
\begin{equation*}
    \boldsymbol{\eta} = 
    \begin{bmatrix}
    \eta_{0}&\hdots&0\\
    \vdots&\ddots&\vdots\\
    0&\hdots&\eta_{W-1}
    \end{bmatrix}_{W\times W},
\end{equation*}
and 
\begin{equation*}
    \mathbf{a}=
    \begin{bmatrix}
        a_{0}^{0}&\hdots&a_{0}^{K}\\
        \vdots&\ddots&\vdots\\
        a_{W-1}^{0}&\hdots&a_{W-1}^{K}
    \end{bmatrix}_{W\times (K+1)}.
\end{equation*}

The coefficients of the polynomial $Q_{K}$ can be retrieved using the annihilation relation in Eq.~(\ref{eq47}). Then by computing the polynomial, we are able to find the 2D locations of projected vertices $z_{k,j}$.

\bibliographystyle{IEEEbib}
%\section*{\refname}
\bibliography{strings, manuscript_ref}

\end{document}